\documentclass[fleqn,usenatbib]{mnras}

\usepackage{newtxtext,newtxmath}

\usepackage[T1]{fontenc}

\DeclareRobustCommand{\VAN}[3]{#2}
\let\VANthebibliography\thebibliography
\def\thebibliography{\DeclareRobustCommand{\VAN}[3]{##3}\VANthebibliography}

\usepackage{graphicx}
\usepackage{xcolor}
\colorlet{BLUE}{blue}

\usepackage{amsmath}	

\title[Time Delay BPL]{Time delay measurements with Broken Power Law model}

\author[Rui et al.]{
Guanhua Rui,$^{1,2,3}$
Bin Hu,$^{1,3}$\thanks{E-mail: bhu@bnu.edu.cn}
Wei Du$^2$\thanks{E-mail: duwei@shnu.edu.cn}
\\
$^{1}$Institute for Frontier in Astronomy and Astrophysics, Beijing Normal University, Beijing, 102206, China\\
$^{2}$Shanghai Key Lab for Astrophysics, Shanghai Normal University, Shanghai, 200234, China\\
$^{3}$School of Physics and Astronomy, Beijing Normal University, Beijing 100875, China
}

\date{Accepted XXX. Received YYY; in original form ZZZ}

\pubyear{2015}
\begin{document}
\label{firstpage}
\pagerange{\pageref{firstpage}--\pageref{lastpage}}
\maketitle


\begin{abstract}
One of the key challenges in strong gravitational lensing cosmography is the accurate measurement of time delays between multiple lensed images, which are essential for constraining the Hubble constant (\(H_0\)). In this study, we investigate how assumptions about the lens mass profile affect time-delay measurements in strong lensing systems. Specifically, we implement a Broken Power Law (BPL) mass model within the \textsc{Lenstronomy} framework \citep{Birrer2018}, which introduces additional flexibility in the radial mass distribution and can phenomenologically capture deviations from a single power-law profile. This model is combined with a numerical approach to compute time delays at the image positions. We validate the BPL implementation using simulated lens systems and compare the results with those obtained from the commonly adopted elliptical power-law (EPL) model. We then apply both model families to the quadruply imaged quasar WGD~2038--4008. 
Both models provide good fits to the imaging and kinematic data overall, with a slight preference for the BPL model. When the internal mass-sheet factor is allowed to vary, the inferred Hubble constant in a flat \(\Lambda\)CDM cosmology with fixed \(\Omega_{\rm m}=0.3\) is \(H_0 = 75.2^{+23.0}_{-16.3} \ \mathrm{km \ s^{-1} \ Mpc^{-1}}\) for the BPL model and \(H_0 = 61.1^{+19.1}_{-13.1} \ \mathrm{km \ s^{-1} \ Mpc^{-1}}\) for the EPL model. For comparison, in the diagnostic case with the internal mass-sheet factor fixed to unity under the same setup, we obtain \(H_0 = 74.2^{+20.3}_{-13.8} \ \mathrm{km \ s^{-1} \ Mpc^{-1}}\) for the BPL model and \(H_0 = 66.1^{+18.8}_{-12.8} \ \mathrm{km \ s^{-1} \ Mpc^{-1}}\) for the EPL model. This highlights how time-delay cosmography remains sensitive to assumptions about the lens mass profile. With current precision, this difference does not favor one cosmological scenario over another, but rather underscores the importance of flexible mass modeling.
\end{abstract}


\section{Introduction}

The Hubble constant (\(H_0\)) characterizes the present-day expansion rate of the Universe, yet its precise value remains one of the most significant unresolved issues in modern cosmology. The so-called Hubble tension refers to the discrepancy, now exceeding the \(5\sigma\) level, between early-universe determinations based on the cosmic microwave background (CMB) and late-universe measurements relying on the distance ladder \citep{Verde2019}. Within the standard \(\Lambda\)CDM framework, CMB observations from the \textit{Planck} satellite imply \(H_0 \simeq 67\,\mathrm{km\,s^{-1}\,Mpc^{-1}}\), in agreement with other early-universe probes \citep{planck2018}. In contrast, direct local measurements yield systematically higher values: the SH0ES collaboration reports \(H_0 = 73.04 \pm 1.04\,\mathrm{km\,s^{-1}\,Mpc^{-1}}\) from Cepheid-calibrated Type~Ia supernovae \citep{riess2022}, with independent methods such as megamaser distances providing consistent results (e.g. \(H_0 = 73.9 \pm 3.0\,\mathrm{km\,s^{-1}\,Mpc^{-1}}\); \citealt{Pesce2020}). Alternative distance-ladder calibrations based on the tip of the red giant branch yield intermediate values around \(H_0 \simeq 69\)--\(70\,\mathrm{km\,s^{-1}\,Mpc^{-1}}\) \citep{Freedman2021}. Meanwhile, inverse distance-ladder and large-scale structure analyses, including baryon acoustic oscillations and galaxy surveys, generally recover \(H_0 \simeq 67\)--\(68\,\mathrm{km\,s^{-1}\,Mpc^{-1}}\), consistent with the CMB results \citep[e.g.,][]{des2018,Addison2018}. The persistence of this tension across multiple, largely independent observational approaches has motivated renewed scrutiny of potential systematic uncertainties and the exploration of extensions beyond the standard \(\Lambda\)CDM model \citep{Verde2019}.

Strong gravitational lensing time-delay cosmography provides an independent, one-step geometric method for measuring the Hubble constant \(H_0\). This technique exploits the relative arrival times between multiple images of a variable background source, such as quasar or supernova, produced by the gravitational potential of a foreground galaxy. By jointly modeling the lens mass distribution and incorporating the source and deflector redshifts, one can infer the time-delay distance \(D_{\Delta t}\), which scales inversely with \(H_0\) for a given cosmological model \citep{Suyu2010,TreuMarshall2016}. As a result, time-delay cosmography does not rely on either local distance ladders or early-universe physics, making it a valuable and complementary probe for addressing the Hubble tension.

Using six lensed quasars with high-quality HST imaging, time-delay measurements, stellar kinematics, and detailed characterization of the lens environment, the H0LiCOW collaboration reported \(H_0 = 73.3^{+1.7}_{-1.8}\,\mathrm{km\,s^{-1}\,Mpc^{-1}}\) in a flat \(\Lambda\)CDM cosmology \citep{Wong2020}. A systematic exploration of modelling systematics and a reanalysis of six time-delay lenses are presented in TDCOSMO~I, yielding \(H_0 = 74.0^{+1.7}_{-1.8}\,\mathrm{km\,s^{-1}\,Mpc^{-1}}\) for composite (stars+halo) models and \(H_0 = 74.2^{+1.6}_{-1.6}\,\mathrm{km\,s^{-1}\,Mpc^{-1}}\) for power-law models \citep{Millon2020TDCOSMOI}. For reference, the STRIDES analysis of DES~J0408$-$5354 reported \(H_0 = 74.2^{+2.7}_{-3.0}\,\mathrm{km\,s^{-1}\,Mpc^{-1}}\) from a single lens \citep{Shajib2020}. Alternative strong-lensing approaches, including free-form mass modeling, have also produced broadly compatible values of \(H_0\), although they likewise highlight the impact of lensing degeneracies and modelling assumptions \citep{Rathna2015,Denzel2021}. On the observational side, in addition to the efforts of the TDCOSMO collaboration, there is a plan to utilize the newly built 1.93-meter aperture optical telescope at the Muztagh-Ata site in the Xinjiang Uygur Autonomous Region of China to monitor lensed quasars and supernovae \citep{Zhu:2022jgv,Dong:2024sij,2025SCPMA..6880407R}. This instrument is expected to improve the accuracy of time delay measurements by approximately a factor of three.

More recent analyses by the TDCOSMO collaboration have focused on quantifying systematic uncertainties associated with lens mass modeling, particularly those related to radial mass-profile assumptions and the mass-sheet degeneracy (MSD).
More recently, multi-plane ray tracing techniques have been applied to gain deeper insights into the line-of-sight (LoS) MSD \citep{2021MNRAS.504.2224L,2021JCAP...08..024F,Lin:2025dgn}.
When the lens mass profile is given increased flexibility and single-aperture stellar kinematics are incorporated with conservative priors, the inferred constraints broaden; for the time-delay lens sample alone, this approach yields \(H_0 = 74.5^{+5.6}_{-6.1}\,\mathrm{km\,s^{-1}\,Mpc^{-1}}\), while including the SLACS--SDSS lenses shifts the result to \(H_0 = 67.4^{+4.1}_{-3.2}\,\mathrm{km\,s^{-1}\,Mpc^{-1}}\) \citep{Birrer2020}. Recent work has highlighted potential systematic issues in SDSS-based stellar velocity dispersions for SLACS lenses; we therefore cite the SLACS-anchored result only for context (see \citealt{Birrer2025}). These results illustrate the sensitivity of time-delay cosmography to lensing mass degeneracies such as the MSD \citep[e.g.,][]{Schneider2013}. In a subsequent blind analysis of the TDCOSMO-2025 sample, incorporating improved stellar kinematics and revised lens modelling, yields \(H_0 = 71.6^{+3.9}_{-3.3}\,\mathrm{km\,s^{-1}\,Mpc^{-1}}\) in a flat \(\Lambda\)CDM cosmology when combined with Pantheon+, with uncertainties that remain intentionally conservative in the treatment of lens-model systematics \citep{Birrer2025}. These results demonstrate that while strong-lensing time-delay cosmography provides competitive constraints on \(H_0\), accurate modeling of the lens mass distribution remains a critical requirement for achieving reliable and precise measurements.

The need for accurate modeling of lens mass distributions in gravitational lensing studies is crucial for obtaining reliable \(H_0\) measurements. Standard models for lens galaxies often assume a power-law radial mass profile (approximately isothermal), but more physically motivated models may be necessary to account for complex processes such as baryonic feedback or self-interacting dark matter (SIDM). SIDM simulations predict that dark matter halos can develop flat-density cores, rather than the steep cusps of cold dark matter (CDM) halos, especially in the central regions of galaxies \citep{TulinYu2018,Kaplinghat2016}. Such core formation (and potential core collapse at later times) can alter the inner mass profile and, in principle, affect lensing observables.

Strong gravitational lensing provides one of the few direct probes of the total mass distribution on kiloparsec and sub-kiloparsec scales, because the positions, flux ratios, and relative time delays of multiple images depend sensitively on the projected lens potential \citep{Sonnenfeld2021,Vegetti2024}. In particular, the presence or absence of a highly de-magnified central image in quad or double lens systems offers a unique constraint on the innermost mass distribution: a steep, cuspy central profile tends to suppress the central image below detectability, whereas a shallower or cored profile can allow such a central image to be brighter and potentially observable \citep{Quinn2016,Perera2023}. Consequently, deep imaging and sensitive radio/mm observations aimed at detecting, or placing stringent upper limits on, demagnified central images have been used in a number of studies to constrain the central density slopes and core sizes of lens galaxies \citep[e.g.,][]{Keeton2003,Quinn2016,Wong2015,Muller2020}. Most such studies report non-detections, which place stringent upper limits on the central-image flux and thereby constrain the allowed core size or, more generally, the allowed central mass distribution. Even in the relatively rare cases of detection, the interpretation is not straightforward: for PMN~J1632$-$0033, \citet{Winn2004} inferred an overall inner slope of \(\sim 1.9\), i.e. close to isothermal, despite the presence of a central image.

A key point is that the central-image flux depends not only on whether the inner galaxy profile is cored or cuspy, but also on the mass of any central supermassive black hole. A central black hole can further demagnify, or even suppress, the central image, so an overall slope inferred to be close to isothermal does not by itself exclude a finite central core or other non-singular inner mass configuration \citep{Du2020}. Conversely, when a central image is detected, its flux can place an upper limit on the black-hole mass and help constrain whether the inner mass distribution must depart from a purely singular profile \citep{Winn2004}. In this sense, a detectable central image generally points to some finite central scale in the mass distribution, whereas a strictly singular central profile would strongly suppress such an image.

Central-image measurements therefore mainly constrain the coupled problem of core size, central surface density, and black-hole mass, rather than providing a direct one-to-one measurement of the smooth inner slope alone. More broadly, these considerations motivate lens models with greater radial flexibility than a single power law. In particular, models that allow the logarithmic slope to vary between the inner and outer regions provide a useful framework for testing whether the mass distribution of lens galaxies remains consistent with a simple cuspy profile or instead exhibits a more complex radial structure. Beyond constraints on the smooth inner mass profile, strong-lensing studies have also been used to probe the nature of dark matter on small scales, including the possible impact of self-interacting dark matter (SIDM) on lens observables \citep{Gilman2021,Gilman2023}. For instance, \citet{Gilman2020_SIDM} demonstrated that flux-ratio anomalies in quadruply imaged quasars can be used to constrain SIDM substructure properties, complementing constraints from dwarf galaxies and galaxy clusters.

In this work, we incorporate the broken power law mass profile \citep[BPL, introduced by][]{Du2020,Du2023} into lens modeling, allowing for a transition in the radial slope of the density profile.  This transition can mimic a core in the inner regions of the lens galaxy. The BPL model offers greater flexibility than the elliptical power law model \citep[EPL, introduced by][]{Barkana1998,Tessore2015} and can phenomenologically reproduce total mass distributions that resemble those predicted in some SIDM scenarios. While the BPL model shares similarities with the cored NFW profile \citep{Navarro1997,Tran2024}, particularly in its ability to describe a central core-like structure, it provides additional flexibility in its radial slope. This flexibility allows the BPL model to capture a broader range of behaviors, from SIDM-like profiles with flat-density cores to the steep gradients typical of CDM halos. As a result, the BPL model serves as a phenomenological and versatile framework for lens modeling, suitable for exploring both SIDM-like core structures and more traditional CDM-like mass distributions. We integrate the BPL model into \textsc{Lenstronomy} \citep{Birrer2018} and develop a numerical solver for time delays, as an analytic Fermat potential is not available in closed form. In Sects.~\ref{lensmass}, we summarize the BPL and EPL mass models and key lensing quantities. Sects.~\ref{tdmsd} validate our approach using mock lensing data, while Sects.~\ref{modeltest} test the BPL model with simulated data. In Sects.~\ref{sec:realimages}, we apply the method to real data of WGD~2038--4008, comparing BPL and EPL model fits. Finally, in Sects.~\ref{sec:results}, we discuss the implications for \(H_0\) inferences and lens model systematics. In Sects.~\ref{app:psidv} we give more detailed explanation to the differences between EPL and BPL.

\section{Lens Mass Models}
\label{lensmass}
This section summarizes the BPL and EPL mass models for gravitational lenses. The BPL profile \citep{Du2020} provides a flexible framework for describing galaxies with central density shallowness or cores, while the EPL model \citep{Barkana1998,Tessore2015} emerges as a special case. All expressions assume an elliptical coordinate defined by the elliptical radius \(R_{\rm el} = \sqrt{q x^2 + y^2/q}\) (with \(q\) the axis ratio) and \(z = x + iy\), ensuring a consistent comparison between models. Key lensing quantities---convergence \(\kappa\), mean convergence \(\bar{\kappa}\), deflection angle \(\boldsymbol{\alpha}\), and deflection potential \(\psi\)---are outlined below for each model, as discussed in Sects.~\ref{sec:mass_profiles},~\ref{sec:maths}, and~\ref{sec:numerical_potential_integration}.

\subsection{Mass Profiles}
\label{sec:mass_profiles}
The BPL mass density profile is defined piecewise with an inner slope \(\alpha_c\) and outer slope \(\alpha\) (with \(\alpha_c < \alpha\) for a core-like center):
\begin{equation}
\kappa(R_{\rm el}) = 
\begin{cases} 
\displaystyle \left( \frac{b}{R_{\rm el}} \right)^{\alpha-1} \left[ \frac{3-\alpha}{2} - \frac{3-\alpha}{\mathcal{B}(\alpha)} \, \mathcal{H}\!\left(\alpha_c, \alpha, \frac{R_{\rm el}}{r_c}\right) \right], & R_{\rm el} \leq r_c, \\[2ex]
\displaystyle \frac{3-\alpha}{2} \left( \frac{b}{R_{\rm el}} \right)^{\alpha-1}, & R_{\rm el} > r_c,
\end{cases}
\end{equation}
where \(\mathcal{B}(\alpha) = \sqrt{\pi}\,\Gamma\!\left(\frac{\alpha-1}{2}\right) / \Gamma\!\left(\frac{\alpha}{2}\right)\), and 
\(\mathcal{H}( \alpha_c,\alpha,\tilde{z}) = \tilde{z} \left[ {}_2F_1\!\left( \frac{\alpha_c}{2}, 1; \frac{3}{2}; \tilde{z}^2 \right) - {}_2F_1\!\left( \frac{\alpha}{2}, 1; \frac{3}{2}; \tilde{z}^2 \right) \right]\)
with \(\tilde{z} = \sqrt{1 - R_{\rm el}^2/r_c^2}\) (here \({}_2F_1\) is the Gaussian hypergeometric function). The parameter \(b\) is a normalization related to the Einstein radius of the lens. The more familiar EPL profile is recovered when \(\alpha_c = \alpha\) or \(r_c \to 0\), which yields 
\begin{equation}
\kappa_{\rm EPL}(R_{\rm el}) = \frac{3 - \alpha}{2} \left( \frac{b}{R_{\rm el}} \right)^{\alpha-1}\,,
\end{equation}
i.e., a single power-law surface density profile \citep{Tessore2015}. In that limit, the inner correction term \(\mathcal{H}\) vanishes and \(b\) directly corresponds to the Einstein radius \(R_{\rm ein}\).

The mean convergence within radius \(R_{\rm el}\) for the BPL model can be derived by integrating \(\kappa\). While the full expression is lengthy \citep[]{Du2020}, an illustrative form is:
\begin{equation}
\bar{\kappa}(<R_{\rm el}) = \frac{1}{\pi R_{\rm el}^2} \int_0^{2\pi}\int_0^{R_{\rm el}} \kappa(R') R' \, dR' d\theta\,,
\end{equation}
which yields, for \(R_{\rm el} \le r_c\),
\begin{equation}
\bar{\kappa} = \left( \frac{b}{R_{\rm el}} \right)^{\alpha-1} + \frac{m_0}{\pi \Sigma_{\rm crit} R_{\rm el}^2} + \frac{2}{3}\frac{3-\alpha}{\mathcal{B}(\alpha)} \left( \frac{b}{r_{\rm c}} \right)^{\alpha-1} \, \mathcal{G}\!\left(\alpha_c, \alpha, \frac{R_{\rm el}}{r_c}\right) ,
\end{equation}
where \(\Sigma_{\rm crit}\) is the critical surface density and \(m_0\) represents a central mass deficit (if \(\alpha_c < \alpha\)) or surplus (if \(\alpha_c > \alpha\)) compared to the extrapolated outer profile. The function \(\mathcal{G}( \alpha_c,\alpha,\tilde{z}) = \tilde{z}^3 \left[ {}_2F_1\!\left( \frac{\alpha}{2}, 1; \frac{5}{2}; \tilde{z}^2 \right) - {}_2F_1\!\left( \frac{\alpha_c}{2}, 1; \frac{5}{2}; \tilde{z}^2 \right) \right]\). For \(R_{\rm el} > r_c\), the last term of \(\bar{\kappa}\) vanishes. In the EPL limit, only the first term remains.

The deflection angle (expressed as a complex quantity \(\alpha^* = \alpha_x - i \alpha_y\)) for the BPL model can be written as \(\alpha^*(z) = \alpha_1^*(z) + \alpha_2^*(z)\), where the first term \(\alpha_1^*\) corresponds to the power-law and the second term \(\alpha_2^*\) is an inner ``compensation'' term due to the break:
\begin{equation}
\alpha_1^*(z) = \frac{R_{\rm el}^2}{z} \left( \frac{b}{R_{\rm el}} \right)^{\alpha-1} {}_2F_1\!\left( \frac{1}{2}, \frac{3-\alpha}{2}; \frac{5-\alpha}{2}; \zeta^2 R_{\rm el}^2 \right),
\end{equation}
with \(\zeta^2 \equiv (q^{-1} - q)/z^2\), and
\begin{equation}
\begin{split}
\alpha_2^*(z) &= \frac{r_c^2}{z} \frac{3-\alpha}{\mathcal{B}(\alpha)} \left( \frac{b}{r_c} \right)^{\alpha-1} \\
&\quad \times \left[ \frac{2}{3-\alpha_c} \, {}_3F_2\!\left(\frac{3-\alpha_c}{2},\frac{1}{2},1; \frac{5-\alpha_c}{2},\frac{3}{2}; \mathcal{C} \right) \right. \\
&\qquad\qquad \left. - \frac{2}{3-\alpha} \, {}_3F_2\!\left(\frac{3-\alpha}{2},\frac{1}{2},1; \frac{5-\alpha}{2},\frac{3}{2}; \mathcal{C} \right) - \mathcal{S}_0 \right] ,
\end{split}
\end{equation}
where \(\mathcal{C} = r_c^2 \zeta^2\), \({}_3F_2\) is a generalized hypergeometric function, and \(\mathcal{S}_0\) is a rapidly convergent series (see \citealt{Du2020} for full definitions). The EPL deflection is recovered by taking \(\alpha_c = \alpha\) (so \(\alpha_2^* \to 0\)) and performing the coordinate transformation for an elliptical power-law potential:
\begin{equation}
\alpha_{\rm EPL}^*(z) = \frac{2b}{1+q} \left( \frac{b}{R_{\rm el}} \right)^{\alpha-1} e^{-i\varphi} \,{}_2F_1\!\left( 1, \frac{\alpha}{2}; 2-\frac{\alpha}{2}; -\frac{1-q}{1+q} e^{-2i\varphi} \right),
\end{equation}
with \(\varphi = \arctan(q x, y)\) the angular coordinate in the lens plane \citep{Tessore2015}. We have verified that \(\alpha_1^*\) for the BPL matches \(\alpha_{\rm EPL}^*\) when \(\alpha_c = \alpha\), as expected.

\subsection{Numerical Potential Integration}
\label{sec:maths}

While deflection angles can be obtained in closed form (in terms of special functions) for the BPL, the deflection potential \(\psi(\mathbf{x})\) generally requires integration of the deflection field. We reconstruct \(\psi\) by numerically integrating \(\boldsymbol{\alpha}(\mathbf{x})\) over the image plane, following the approach of \citet{Keeton2001B}. For completeness, we outline the integration procedure in an elliptical coordinate system. 
 
We adopt coordinates \((x,y)\) aligned with the lens principal axes, and define an elliptical radius 
\begin{equation}
\xi^2 = q x^2 + \frac{y^2}{q}\,,
\end{equation}
which differs from \(R_{\rm el}^2\) by a constant factor for fixed \(\xi\). We introduce a parameter \(u \in [0,1]\) such that 
\begin{equation}
\xi(u)^2 = u \left( q x^2 + \frac{y^2}{1 - (1 - q^2)u} \right)\,,
\end{equation}
which smoothly interpolates \(\xi\) from 0 to \(\infty\) as \(u\) goes from 0 to 1 \citep[see][]{Schramm1990}. Via 2D Green function method, the surface mass density can be expressed as \(\kappa(\xi)\) and the potential at \(\mathbf{x}\) is 
\begin{equation}
\psi(\mathbf{x}) = \frac{1}{\pi} \int \kappa(\xi(\mathbf{y})) \ln|\mathbf{x}-\mathbf{y}| \, d^2y\,.
\end{equation}
Performing the angular average and changing variables from area element \(d^2y\) to \(du\), one obtains 
\begin{equation}
\psi(x,y) = \frac{1}{2} \int_0^1 \frac{\xi(u)}{u\,\sqrt{1 - (1 - q^2)u}}\, \phi_r(\xi(u))\, du\,,
\end{equation}
where 
\begin{equation}
\phi_r(\xi) = \frac{2}{\xi} \int_0^{\xi} s\,\kappa(s)\,ds 
\end{equation}
is the circularly symmetric potential for surface density \(\kappa(s)\) evaluated at \(\xi\). We compute \(\phi_r(\xi)\) for the BPL profile by integrating the \(\kappa(R_{\rm el})\) expression (treating \(\xi \approx R_{\rm el}\) for the circular case), and then perform the \(u\)-integration numerically. The result is the deflection potential \(\psi(x,y)\) up to an arbitrary additive constant (which does not affect relative time delays). 

We verified this integration procedure by applying it to the EPL model, for which \(\psi_{\rm EPL}(x,y)\) is known analytically. Specifically, for an elliptical power-law lens one has \(\psi_{\rm EPL}(\mathbf{x}) = \frac{1}{2-\alpha} (x \alpha_x + y \alpha_y)\) \citep{Tessore2015}.

\subsection{Validation of Fermat Potential Solver}
\label{sec:numerical_potential_integration}
Theoretically, the integration method described above should be applicable to any mass profile with elliptical symmetry. We first validate our numerical deflection potential solver using the EPL model, which has an analytic form for \(\psi\). To test our method, we compute Fermat potential values on a grid in the image plane both analytically and via numerical integration, then compare the two. The fractional difference at each grid point is given by:

\begin{equation}
{\rm Res}(x,y) = \frac{|\psi_{\rm int}(x,y) - \psi_{\rm ana}(x,y)|}{|\psi_{\rm ana}(x,y)|}\,,
\end{equation}
where \(\psi_{\rm int}\) is the integrated potential and \(\psi_{\rm ana}\) is the exact solution. 

\begin{figure}
    \centering
    \includegraphics[width=\columnwidth]{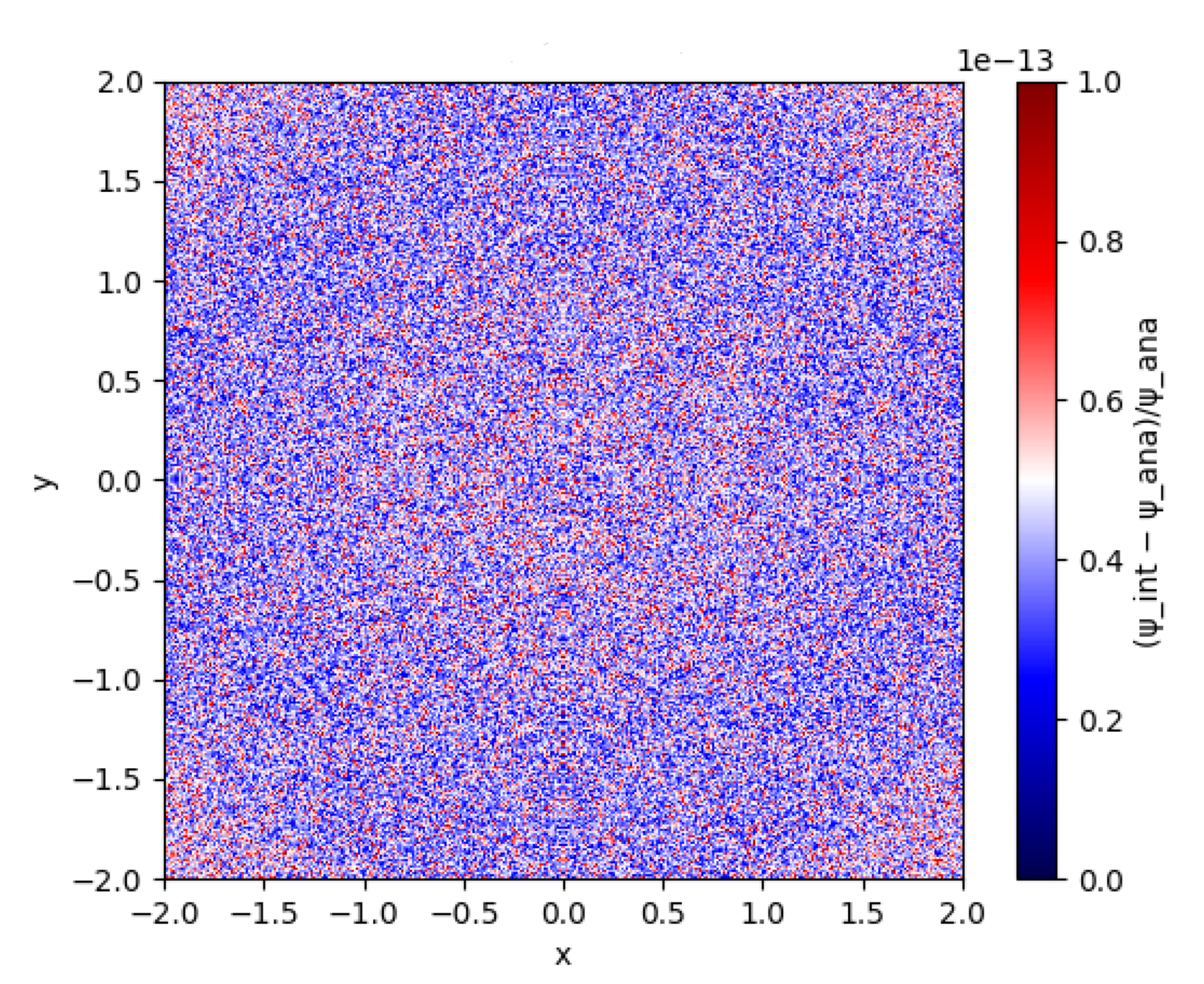}
    \caption{Relative residual between the numerically integrated lens potential and the analytic solution, for an EPL model. We show the relative residual \({\rm Res}(x,y)\) across the image plane for a test lens (Einstein radius \(\theta_E = 1\farcs5\), slope \(\alpha=1.9\), axis ratio \(q=0.8\)). The residuals are below \(10^{-12}\) everywhere, demonstrating the high accuracy of the integration scheme.}
    \label{fig:image1}
\end{figure}

Fig.~\ref{fig:image1} shows a map of the residuals for an example power-law lens (with parameters \(\theta_E = 1\farcs5\), \(\alpha=1.9\), \(q=0.8\), oriented along the axes). The residuals are \(\lesssim 10^{-12}\) without systematic pattern, confirming that the integration yields the correct potential to machine precision. This level of accuracy far exceeds the requirements for time-delay calculations, which depend on potential differences between image points. Moreover, evaluating \(\psi\) at a few image positions, rather than over a full grid, is highly efficient. We conclude that our numerical approach to computing the Fermat potential is both accurate and efficient, and will be used to evaluate time delays for lenses modeled with the BPL profile.

\section{Time Delay and Mass-Sheet Degeneracy}
\label{tdmsd}
This section introduces the time-delay relation for gravitational lenses and discusses the impact of the MSD on time-delay measurements. The time-delay difference between two images in a gravitational lens system depends on the mass distribution, the deflector's geometry, and the cosmological parameters, as described in Sect.~\ref{sec:time_delay_relation}. To constrain the degeneracy and improve the precision of Hubble constant measurements, we incorporate stellar kinematics and external convergence measurements into the lens model, as outlined in Sects.~\ref{sec:kinematics_analysis} and~\ref{sec:mass_sheet_degeneracy}.

\subsection{From Time Delay to Hubble Constant}
\label{sec:time_delay_relation}
The time delay \(\Delta t_{XY}\) between the arrival times of photons corresponding to images X and Y is given by the following expression:
\begin{equation}
\Delta t_{XY} = \frac{D_{\Delta t}}{c} \left( \phi(\theta_X) - \phi(\theta_Y) \right),
\end{equation}
where \(D_{\Delta t}\) is the time-delay distance, \(c\) is the speed of light, and \(\phi(\theta)\) is the Fermat potential, defined as:
\begin{equation}
\phi(\theta) = \frac{1}{2} |\theta - \beta|^2 - \psi(\theta),
\end{equation}
with \(\theta\) being the angular position of the image, \(\beta\) the angular position of the source, and \(\psi(\theta)\) the lens potential.

The time-delay distance \(D_{\Delta t}\) is related to the Hubble constant \(H_0\) by the relation:
\begin{equation}
D_{\Delta t} \propto H_0^{-1}.
\end{equation}
Thus, time delays can be used to measure \(H_0\) by determining the time-delay distance, which is inversely proportional to \(H_0\).

\subsection{Mass-Sheet Degeneracy}
\label{sec:kinematics_analysis}
The imaging observables of the lensing phenomenon—such as the image positions and flux ratios\footnote{The absolute flux value is affected by the MST.}—remain invariant under the transformation known as the mass-sheet transformation (MST) \citep{Falco1985}, which rescales the convergence as:
\begin{equation}
\kappa(\theta) \to \kappa_{\lambda}(\theta) = \lambda \kappa(\theta) + 1 - \lambda,
\end{equation}
where \(\lambda\) is a rescaling factor. 

We can decompose the total mass convergence \(\kappa_{\rm true}\) into two components: the central deflector’s convergence \(\kappa_{\rm cen}\) and the external convergence \(\kappa_{\rm ext}\), which arises from the mass distribution along the line-of-sight (LOS). The true convergence is then:
\begin{equation}
\kappa_{\rm true} = \kappa_{\rm cen} + \kappa_{\rm ext}.
\end{equation}
The model convergence \(\kappa_{\rm model}\) is related to the true convergence \(\kappa_{\rm true}\) by the mass-sheet transformation as \citep{Birrer2016}:
\begin{equation}
\kappa_{\rm model} = \frac{\kappa_{\rm true} - \kappa_{\rm ext}}{1 - \kappa_{\rm ext}}.
\end{equation}
However, the lens model \(\kappa_{\rm model}\) that we actually constrain can be an internal MST of \(\kappa_{\rm model}\), as:
\begin{equation}
\kappa_{\rm model}^\prime = \lambda_{\rm int} \kappa_{\rm model} + 1 - \lambda_{\rm int},
\end{equation}
where \(\lambda_{\rm int}\) is the internal rescaling factor. This equation shows that the lens model we constrain can differ from the true convergence by an internal rescaling factor \(\lambda_{\rm int}\). The physical meaning of the internal mass sheet is originated from the cored profile of the main lenses\citep[e.g.,][]{Schneider2013,2021A&A...652A...7C,Blum2020}. 
For internal mass-sheet degeneracy, the true mass distribution \(\kappa_{\rm true}\) is then related to the internally rescaled model mass distribution \(\kappa_{\rm model}\) as:
\begin{equation}
\kappa_{\rm true} = (1 - \kappa_{\rm ext}) [\lambda_{\rm int} \kappa_{\rm model} + 1 - \lambda_{\rm int}] + \kappa_{\rm ext}.
\end{equation}
Luckily, it can be determined by fitting the model to other constraints, such as stellar kinematics.

This equation shows how the true convergence \(\kappa_{\rm true}\) can be recovered from the model convergence \(\kappa_{\rm model}\) by using the external convergence \(\kappa_{\rm ext}\) and the internal scaling factor \(\lambda_{\rm int}\). Importantly, when \(\kappa_{\rm ext}\) is independently known, the lens model can be corrected by inverting the MST. However, if \(\kappa_{\rm ext}\) is not known, this introduces a degeneracy in the time delay and other lensing observables.

The time delay is rescaled with respect to the mass-sheet transformation as:
\begin{equation}
\Delta t \to \Delta t' = \Lambda_{\rm tot} \Delta t.
\end{equation}
In the presence of both internal and external rescaling, the true time delay is related to the model time delay \(\Delta t_{\rm model}\) by:
\begin{equation}
\Delta t_{\rm true} = \Lambda_{\rm tot} \Delta t_{\rm model} = (1 - \kappa_{\rm ext}) \lambda_{\rm int} \Delta t_{\rm model}.
\label{eq:dttrue}
\end{equation}
Thus, to constrain the degeneracy, it is crucial to obtain additional constraints, such as stellar kinematics, which allow us to estimate \(\lambda_{\rm int}\) and LoS weak lensing or galaxy number counts to estimate \(\kappa_{\rm ext}\). By incorporating these extra measurements, the degeneracy can be broken, and the true mass distribution, including the correct time delay, can be determined.

\subsection{Kinematics Analysis}
\label{sec:mass_sheet_degeneracy}

The stellar velocity dispersion, \(\sigma_{\rm los}\), constrains the mass distribution of the deflector galaxy. The kinematics are governed by the Jeans equation:
\begin{equation}
\frac{d}{dr} \left( l(r) \sigma_r(r)^2 \right) + 2 \beta_{\rm ani}(r) \frac{l(r) \sigma_r(r)^2}{r} = -l(r) \frac{d\Phi(r)}{dr},
\end{equation}
where \(l(r)\) is the 3D luminosity density, \(\sigma_r(r)\) is the radial velocity dispersion, and \(\beta_{\rm ani}(r) = r^2/(r^2 + r_{\mathrm{ani}}^2)\) is the Osipkov-Merritt (OM) anisotropy parameter \citep[]{Osipkov1979,Merritt1985}.

A constant anisotropy parametrization is a viable alternative and has been adopted in a number of dynamical analyses. In the present work, however, we retain the OM form to remain consistent with our current modelling setup and with previous analyses of WGD~2038--4008, thereby enabling a controlled comparison with earlier results. For a single-aperture velocity-dispersion measurement, such a one-parameter, physically motivated prescription provides a pragmatic way to marginalize over orbital anisotropy without introducing a level of flexibility that is not supported by the data. More generally, modern Jeans-based dynamical modelling has emphasized both the importance of anisotropy assumptions and the feasibility of adopting more general anisotropy prescriptions (e.g., \citealt{Cappellari2026}); we therefore treat the anisotropy model as a potential source of systematic uncertainty.

The observable line-of-sight velocity dispersion, \(\sigma_{\rm los}(R)\), is related to the 3D velocity dispersion by:
\begin{equation}
\label{eq:losmodel}
\sigma_{\rm los}^2(R) = \frac{2 G}{I(R)} \int_R^{\infty} K_{\beta} \left( \frac{r}{R} \right) l(r) M(r) \frac{dr}{r},
\end{equation}
where \(I(R)\) is the surface brightness, \(K_{\beta}\) is a function of \(\beta_{\rm ani}\), and \(M(r)\) is the 3D enclosed mass \citep{Mamon2005}. The corresponding form of \(K_{\beta}(u \equiv r/R)\) is:
\begin{multline}
\mathcal{K}_\beta = \frac{u_{\mathrm{ani}}^2 + 1/2}{(u_{\mathrm{ani}} + 1)^{3/2}} \left( \frac{u^2 + u_{\mathrm{ani}}^2}{u} \right) \tan^{-1} \left( \sqrt{\frac{u^2 - 1}{u_{\mathrm{ani}}^2 + 1}} \right) \\
- \frac{1/2}{u_{\mathrm{ani}}^2 + 1} \sqrt{1 - \frac{1}{u^2}},
\end{multline}
where \(u_{\mathrm{ani}} = r_{\mathrm{ani}}/R\). The observed aperture-averaged velocity dispersion is:
\begin{equation}
\label{eq:sigmav_model}
\sigma_{\mathrm{ap}}^2 = \frac{\int_{\mathrm{ap}} \left[I(R) \sigma_{\mathrm{los}}^2(R)\right] * \mathcal{S} \, dx \, dy}{\int_{\mathrm{ap}} I(R) * \mathcal{S} \, dx \, dy},
\end{equation}
where \(* \mathcal{S}\) denotes convolution with the point spread function (PSF). The lens-model-predicted LOS velocity dispersion is:
\begin{equation}
\sigma_{\mathrm{ap}, \text{model}}^2 = \frac{D_{\mathrm{s}}}{D_{\mathrm{ds}}} c^2 J\left(\xi_{\mathrm{lens}}, \xi_{\mathrm{light}}, \beta_{\mathrm{ani}}\right).
\end{equation}
Finally, accounting for internal and external mass-sheet transformations:
\begin{equation}
\label{eq:sigmav_combine}
\sigma_{\mathrm{ap}, \text{true}}^2 = \left(1 - \kappa_{\mathrm{ext}}\right) \lambda_{\mathrm{int}} \sigma_{\mathrm{ap}, \text{model}}^2.
\end{equation}
This rescaling constrains the MSD by matching the observed and model-predicted velocity dispersions. By measuring the velocity dispersion and external convergence, we can correct the time delay for the MST and recover the true mass distribution, leading to unbiased cosmological measurements, including the Hubble constant.

\section{Model Tests on Simulated Lenses}
\label{modeltest}

We now assess the performance of the BPL model using simulated lens systems. Our objectives are twofold: (i) to evaluate if fitting a lens with a true BPL profile using an EPL model introduces biases in image-only modeling, as discussed in Sect.~\ref{sec:ep_fits_bpl_simulated_data}, and (ii) to confirm that fitting a BPL model to BPL-simulated data accurately recovers the input parameters, as outlined in Sect.~\ref{sec:bpl_fits_bpl_simulated_data}, thus validating our implementation. These tests will provide insight into the potential systematic errors introduced when a simplified mass model, such as the EPL, is used for lenses with a steeper inner density slope.

\subsection{EPL fits to BPL-simulated data}
\label{sec:ep_fits_bpl_simulated_data}
First, we simulate lensing data from a BPL mass model and fit it using a standard EPL model. We choose BPL parameters informed by cores discussed in \citet{Du2020}. Specifically, we take an outer slope \(\alpha=2.0\) (nearly isothermal), an inner slope \(\alpha_c=0.5\) (significantly flattened core), and a break radius \(r_c = 0\farcs5\). We set the axis ratio \(q=0.8\). To isolate the effect, we set the external shear to zero in this test. We then generate a mock lens image with a quadruply imaged background point source plus extended host galaxy emission, adding a smooth lens light profile. In both the simulation and the fitting, we adopt a common, idealized Gaussian PSF with a full width at half maximum (FWHM) of \(0\farcs1\).

\begin{figure}
    \centering
    \includegraphics[width=0.9\columnwidth]{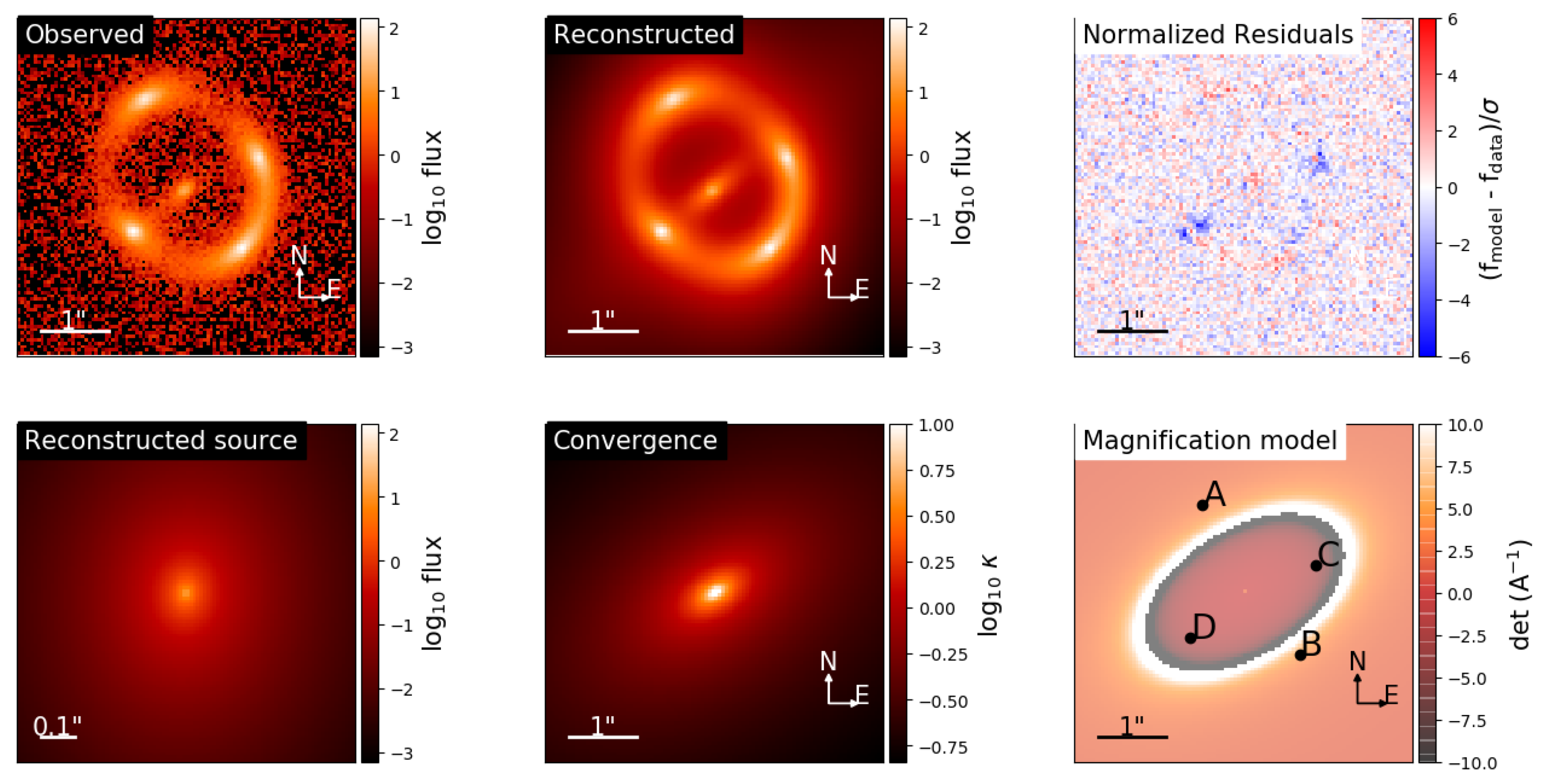}
    \caption{Validation test: Fitting EPL model to the BPL mock data. \textit{Top Left:} Simulated data with four quasar images and host galaxy arcs. \textit{Top Middle:} Best-fitting EPL model reconstruction. \textit{Top Right:} Normalized residuals. \textit{Bottom Left:} Reconstructed source in the source plane. \textit{Bottom Middle:} Convergence map of the best-fit EPL model, showing an elliptical mass distribution. \textit{Bottom Right:} Magnification map with the four quasar image positions (A--D) labeled. We assume an idealized Gaussian PSF with \(\mathrm{FWHM}=0\farcs1\) in both the simulation and the fitting. The best-fit EPL reconstruction yields a reduced chi-square of \(\chi^2 \simeq 1.08\).
    }
    \label{fig:image2}
\end{figure}

We fit the mock data with an EPL mass distribution, jointly modeling the lens galaxy light (\text{S\'{e}rsic} profile), the lensed point source, and the extended host arcs (\text{S\'{e}rsic} profile). The results are shown in Fig.~\ref{fig:image2}. Even with lens light included, the EPL model failed to fully reproduce the lensed host arcs: coherent residuals remain near the image positions. This behavior is expected because the true mass distribution contains a core (shallower central \(\kappa\)), whereas the single–power-law EPL compensates by adjusting its global slope and ellipticity. We also note that the bright lens galaxy, overlapping the faint host, can partially mask mass-model deficiencies: the fitting algorithm may absorb some flux mismatches into the lens-light component. Consequently, relatively small residuals at the point-image locations should not be over-interpreted as evidence for an adequate mass model—the light model can soak up discrepancies that originate in the mass profile.

\subsection{BPL fits to BPL-simulated data}
\label{sec:bpl_fits_bpl_simulated_data}
\begin{figure}
    \centering
    \includegraphics[width=0.9\columnwidth]{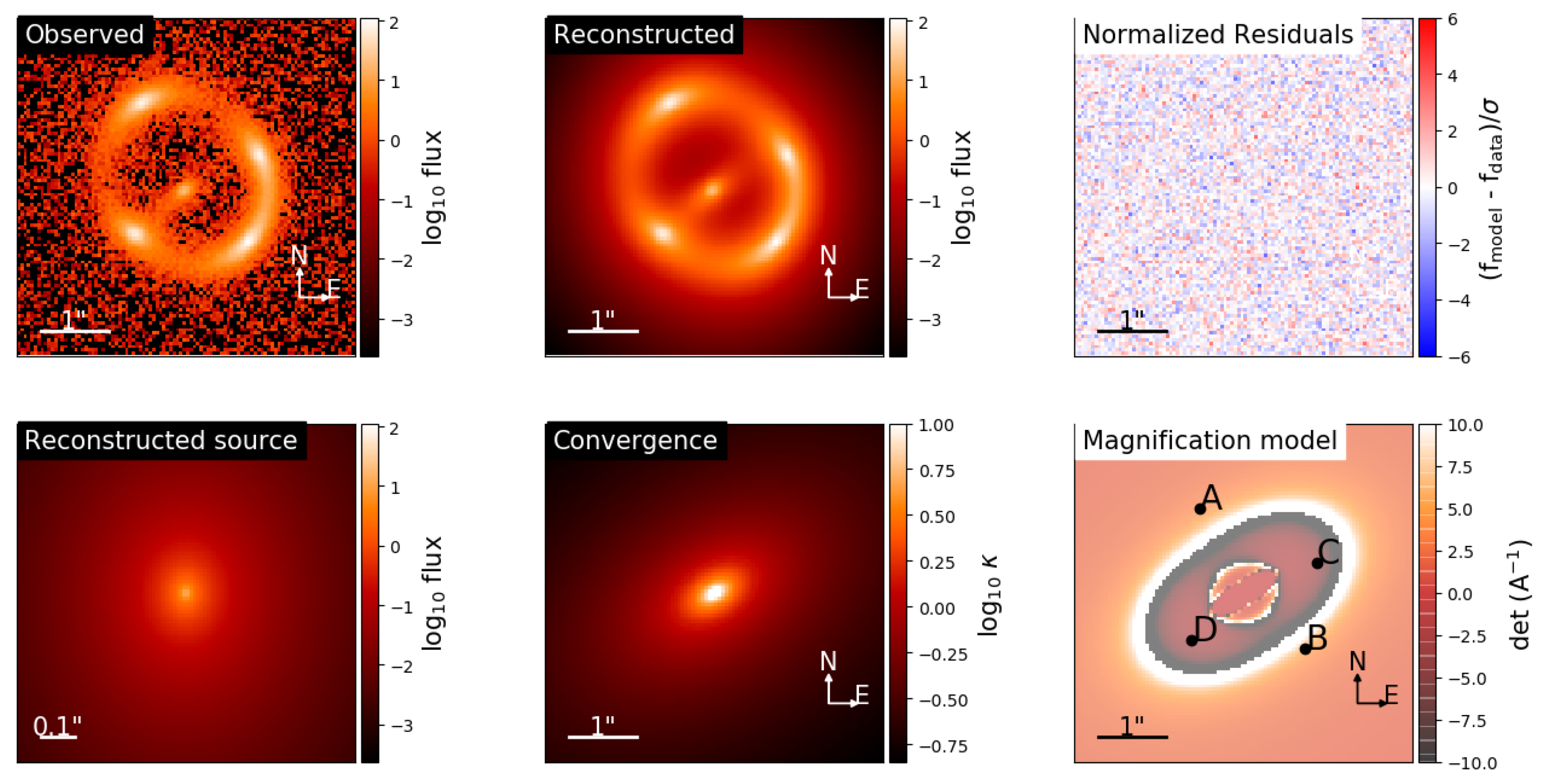}
    \caption{Validation test: Fitting BPL model to the BPL mock data. Panels are analogous to Fig.~\ref{fig:image2}. The BPL model successfully reproduces the lensed images and arcs, yielding negligible residuals. The input and recovered mass profile parameters agree within uncertainties. The best-fit BPL reconstruction yields a reduced chi-square of \(\chi^2 \simeq 1.00\).
    }
    \label{fig:image3}
\end{figure}

Next, to verify the internal consistency of our BPL implementation, we simulate lens data using a BPL model and then fit it with the same functional form. We use the same parameters as above (outer slope \(\alpha=2.0\), inner \(\alpha_c=0.5\), \(r_c=0.5\), etc.) to generate mock images, and then run our BPL fitting pipeline. As shown in Fig.~\ref{fig:image3}, the BPL model provides an excellent fit to the data, with residuals consistent with noise. The recovered mass parameters are within the statistical uncertainties of the true input values. This confirms that: (i) the simulation and fitting procedures are properly coded and inverses of each other, and (ii) the BPL model is flexible enough to capture the lensing observables when it is indeed the correct model. 

These simulation tests are designed to validate the internal consistency of our imaging-modeling implementation and to illustrate the impact of mass-model mismatch in recovering the lens mass distribution from imaging data. Accordingly, any differences in the predicted time delays discussed later are most naturally interpreted as arising, at least in part, from how such mass-model mismatch propagates into the Fermat potential.

\section{Lens Modeling of WGD~2038--4008}
\label{sec:realimages}

We now apply our modeling approach to the quadruply imaged quasar WGD~2038--4008. This system was discovered in the Dark Energy Survey footprint \citep{Agnello2018} and consists of a lens galaxy at \(z_{\ell}=0.228\) and a background quasar at \(z_{s}=0.777\) \citep{BuckleyGeer2020}. It has four resolved quasar images (labeled A–D) and faint host galaxy arcs visible in deep imaging. Crucially, a stellar velocity dispersion of \(296 \pm 19\) km~s\(^{-1}\) (aperture size \(0\farcs75 \times 1\farcs0\), seeing \(0\farcs9\)) has been measured for the lens galaxy using Gemini/GMOS-S spectroscopy \citep{BuckleyGeer2020}, providing an additional constraint for our mass models. This data set makes WGD~2038--4008 an excellent candidate for time-delay cosmography. The choice of this lens system is motivated by the fact that, in previous fits, the system has shown potential for a central imaging feature \citep{Shajib2019,Shajib2022}.

Our modeling procedure closely follows that of recent lensing analyses \citep[e.g.,][]{Wong2020,Shajib2020}. We describe the imaging data and preparation in Sect.~\ref{subsec:data}, then define two model settings in Sect.~\ref{subsec:models}. We perform Bayesian inference for each family, including an exploration of model systematics and a model averaging procedure in Sect.~\ref{subsec:bic}. Finally, in Sect.~\ref{comlos}, we combine stellar kinematics and external convergence to refine the lens model and constrain the mass-sheet factor, enabling more precise time-delay predictions and cosmological inferences.

\subsection{Data and Preprocessing}
\label{subsec:data}

We use \textit{HST} imaging of WGD~2038--4008 from the WFC3 camera in three filters: F160W (near-infrared), F814W (optical \(I\)-band), and F475X (blue). Each filter has four dithered exposures (total exposure times of 2280~s in F160W, 2400~s in F814W, 2268~s in F475X), which we retrieved from the Mikulski Archive for Space Telescopes. The data were processed with standard \textsc{DrizzlePac} routines to produce combined images in each band with a pixel scale of \(0\farcs08\) for WFC3/IR and \(0\farcs04\) for WFC3/UVIS, respectively. We constructed empirical PSFs for each band by stacking 4--6 unsaturated stars in the field, and then iteratively refined the PSF models using the quasar image residuals \citep[e.g., following][]{Chen2016}. The images were astrometrically aligned to sub-pixel precision using common sources across the bands. A similar image download and preprocessing procedure, from which our pipeline draws inspiration, can be found in \cite{Tan2024}\footnote{\url{https://github.com/Project-Dinos/dinos-i}}, although it was not specifically designed for this target.

\begin{figure}
    \centering
    \includegraphics[width=\columnwidth]{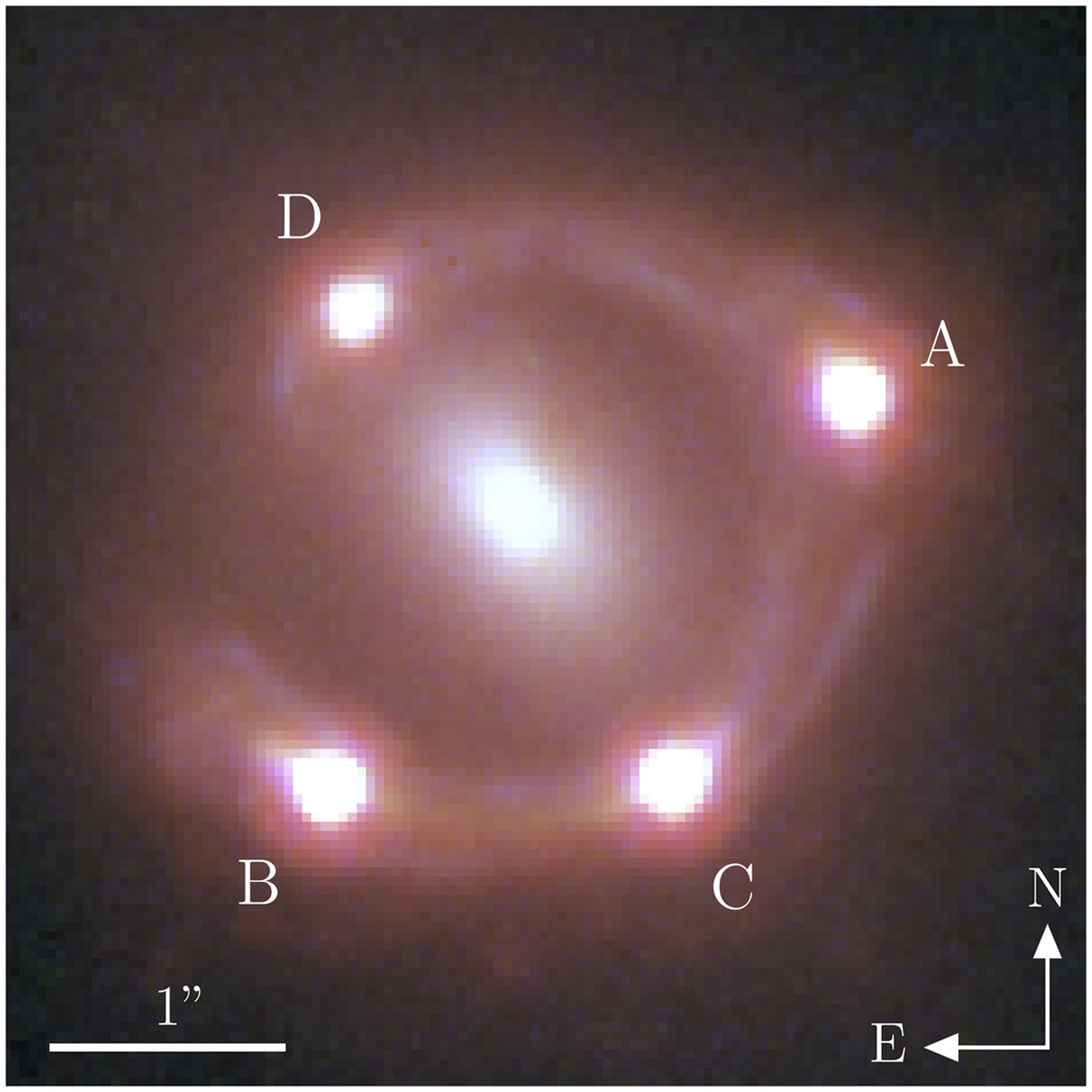}
    \caption{Color composite of WGD~2038--4008 from \textit{HST} WFC3 imaging (F160W in red, F814W in green, F475X in blue). The four quasar images A, B, C, D surround the foreground lens galaxy (center). Extended lensed host galaxy features are visible as faint greenish arcs. North is up and east is left; the image cutout is \(5''\) on a side (see also \citealt{Wong2024}).}
    \label{fig:wgd_rgb}
\end{figure}

Fig.~\ref{fig:wgd_rgb} shows a false-color composite of the system (see also \citealt{Wong2024}). The four quasar images (A--D) form a classic fold/cusp configuration around the lens galaxy. The quasar is relatively red, appearing brightest in F160W; the extended host galaxy is detected mainly in F160W and F814W (as greenish arcs), with marginal signal in F475X. We mask out neighboring objects and regions without detected lensing signal when fitting the pixel data. The four exposures in each filter were designed to cover the large range of light intensity between the bright quasar images and the much fainter host-galaxy arcs. The final science images in each band are produced by combining the dithered exposures with the standard HST reduction and \textsc{DrizzlePac} pipeline. In the lens modelling we perform a joint fit to the combined images in F160W, F814W, and F475X, while masking unrelated companions and regions without detected lensing signal.

\subsection{Model Settings and Priors}
\label{subsec:models}

We model the lens with two closely related mass-profile families-EPL(PEMD model) and BPL-as discussed and compared above, while keeping the deflector light model fixed to the deflector configuration, so that mass–light coupling and PSF handling remain fully consistent. Imaging constraints are taken from HST WFC3 F160W, F814W, and F475X. During optimization we iteratively reconstruct the PSF around the quasar images, and we restrict the image-plane likelihood to a circular aperture of radius \(2\farcs2\) for the WFC3/IR image and \(3\farcs6\) for the two UVIS images, masking unrelated companions within these regions. \citet{Shajib2022} used slightly larger masks of \(2\farcs3\) in IR and \(3\farcs7\) in UVIS. However, we find that this choice leads to a poorer BIC value, so we adopt the smaller apertures in the present analysis.

To account for residual PSF-modelling systematics in the imaging likelihood, we follow \citet{Shajib2022} and inflate the pixel variance map in the vicinity of the quasar point-source regions by adding a PSF-uncertainty term. The PSF-uncertainty map is constructed from the initial PSF estimate and effectively corresponds to adding a small fractional flux uncertainty to the variance map around the quasar images.

Under a shared elliptical isodensity geometry, both families allow the mass centroid \((\theta_1, \theta_2)\) to vary independently of the light centroid and include an external shear \((\gamma_{\rm ext}, \phi_{\rm ext})\). The EPL is characterized by a single Einstein radius \(\theta_{\rm E}\), an axis ratio \(q_{\rm m}\) with position angle \(\phi_{\rm m}\), and a single slope \(\gamma\). The BPL retains the same \((q_{\rm m}, \phi_{\rm m})\) but replaces the single slope with an inner/outer pair \((\alpha_c, \alpha)\) joined at a break radius \(r_c\), thereby capturing curvature in the radial mass profile while preserving the overall ellipticity and centroid. Additionally, the BPL model includes a scale radius \(b\) corresponding to \(\theta_{\rm E}\). Because the Einstein radius of the BPL model cannot be expressed analytically, we adopt the effective Einstein radius \( \theta_{\rm E}\) in this work. The effective Einstein radius is computed as the circularized radius \( \theta_{\rm E}\) at which the enclosed mean convergence satisfies \(\bar{\kappa}(<\theta_{\rm E}) = 1\) (i.e.\ defined via circular, not elliptical, apertures).

The source is reconstructed using a compact basis, where both the elliptical \text{S\'{e}rsic} and shapelets \citep{Refregier2003,Birrer2015} components share the same coordinate system. The reconstruction is performed jointly across multiple bands to ensure consistency. For the shapelets model, the number of shapelet modes \(n_{\max}\) is set based on the resolution requirements for each band. Specifically, \(n_{\max}\) defines the maximum order of the shapelet decomposition, controlling the number of radial and angular modes used to describe the source profile. A higher \(n_{\max}\) allows for a finer resolution but also increases the complexity of the model. For the shapelets decomposition is parametrized by the maximum order \(n_{\max}\), with the choices \(\{n^{IR}
_{max} , n^{UVIS}_{max} \} = \{7, 11\},\{8, 12\}\) and \(\{9, 13\}\). The deflector light is described by a triple \text{S\'{e}rsic} model (three \text{S\'{e}rsic} share the same coordinates) that permits colour gradients by freeing $(I_{\rm eff},\theta_{\rm eff},n_{\rm s})$ per band while tying the centroids across bands, so that differences between EPL and BPL arise from the mass model rather than the light parameterization. The specific analytical expressions for the light profile can be found in Appendix~\ref{app:light}.

Priors are weakly informative and follow standard cosmographic practice: the mass centroid is uniform within \(\pm0\farcs2\) of the light centroid; \(\theta_{\rm E}\) is uniform over a plausible interval bracketing the isophotal Einstein scale; \(q_{\rm m}\in[0,1]\) and \(\phi_{\rm m}\in[0,180^\circ)\); for EPL we take \(\gamma\in[1.5,2.8]\); for BPL we adopt \(\alpha\in[1.0,3.0]\), \(\alpha_c\in[0.01,3.0]\) with a soft regularization that enforces monotonic enclosed mass and \(r_c\in[0\farcs04,3\farcs0]\); for the tidal field, \(\gamma_{\rm ext}\in[0,0.2]\) and \(\phi_{\rm ext}\in[0,180^\circ)\). Light parameters use broad uniforms (or weak Gaussians) that ensure positive fluxes and realistic sizes/indices.

For each model configuration, we first perform particle swarm optimization (PSO) followed by iterative PSF updates using the \texttt{psf\_iteration} function in \textsc{lenstronomy}. After each PSF update, we re-align the IR frame to the UVIS frame using the quasar image positions and then re-optimize the model. This process alternates between PSO and PSF iterations, ensuring the model is progressively refined. Once the initial optimization is completed, we perform MCMC with 10,000 burn-in steps followed by 10,000 MCMC steps (Run 1). Afterward, we repeat the burn-in with 10,000 steps and MCMC with 10,000 steps (Run 2) to account for the influence of the iterative PSF updates. Convergence is monitored by verifying that the walker medians and dispersions have stabilized after 1,000 steps, and the final 1,000 steps are used for posterior analysis.

The complete modelling workflow, including all testing and verification runs, accumulated a total computational cost of approximately \(10^5\) core-hours. In fact, due to the BPL model involving more complex calculations of confluent hypergeometric functions and the need for integration in the lens potential calculation, the overall computation took approximately 1/3 to 1/2 more core-hours compared to the EPL model. For the EPL model family, the overall model configuration and optimization strategy closely follow the publicly available implementation described in \citet{Shajib2022}, as documented in the accompanying GitHub repository\footnote{\url{https://github.com/TDCOSMO/WGD2038-4008}}. This ensures that our EPL results are directly comparable to previous analyses of the same system. The full set of model configurations adopted in this work, together with the corresponding posterior samples for both the EPL and BPL families, will be made publicly available in a dedicated repository upon publication.

\subsection{Bayesian Information Criterion}
\label{subsec:bic}

We compare different modeling configurations using the Bayesian Information Criterion (BIC). The BIC is not used here as an absolute model-selection statistic, since the imaging likelihood is evaluated over a large number of highly correlated pixels and therefore does not strictly satisfy the independence assumptions underlying the BIC. Instead, in this work the BIC is employed only as a relative weighting tool within a controlled model family (EPL or BPL separately), where all configurations share the same data set, likelihood definition, masking strategy, and PSF treatment. In this context, the role of the BIC is to stabilize the ranking of closely related model realizations against numerical fluctuations arising from finite optimization accuracy and sampling noise, rather than to provide a rigorous estimate of Bayesian evidence. Consequently, no absolute evidence comparison between the EPL and BPL model families is attempted, and the resulting weights are not interpreted as a statement of global model preference. For each MCMC chain (i.e., each modeling configuration), we compute the BIC from the maximized image-likelihood. The BIC is defined as:

\begin{equation}
\mathrm{BIC} \;=\; k \,\ln n \;-\; 2 \ln \hat{\mathcal L},
\end{equation}
where \(n\) is the number of independent data points used by the likelihood. In our imaging case, \(n\) corresponds to the total number of unmasked pixels across all bands and exposures. \(k\) is the total number of fitted parameters (both non-linear and linear), and \(\hat{\mathcal L}\) is the maximum value of the image likelihood attained by the chain. We compute:
\begin{equation}
k \;=\; k_{\rm nonlin} \;+\; k_{\rm lin},
\end{equation}
where \(k_{\rm nonlin}\) is the number of non-linear parameters in all model components (lens mass, source light, lens light, point sources, etc.), and \(k_{\rm lin}\) is the number of linear amplitudes solved at each likelihood evaluation (i.e., linear light coefficients).

We calculate one \(\mathrm{BIC}\) value for each sampling chain. Within each model family, we evaluate the relative performance of different parameter sets by computing the \(\Delta\mathrm{BIC}\) values as
\begin{equation}
\Delta \mathrm{BIC}_i \;=\; \mathrm{BIC}_i \;-\; \min_j(\mathrm{BIC}_j),
\end{equation}
where both \(\mathrm{BIC}_i\) and \(\min_j(\mathrm{BIC}_j)\) are evaluated within the same model family, and the index \(j\) corresponds to the parameter set that yields the minimum \(\mathrm{BIC}\) in that family. This definition allows us to measure how strongly each realization is disfavored relative to the best-fitting model of its family. And the classical BIC weights are calculated as:
\begin{equation}
w_i \propto \exp\!\left(-\tfrac{1}{2}\Delta \mathrm{BIC}_i\right).
\end{equation}
However, finite optimization and sampling introduce scatter in the BIC values. To account for this, we adopt a robust weighting scheme that marginalizes over an effective BIC noise scale. We define \(\sigma_{\rm num}\) as the numerical scatter (estimated from repeated runs of identical configurations) using a robust median absolute deviation (MAD) estimator, and \(\sigma_{\rm mod}\) as the typical inter-configuration variation (estimated from differences between adjacent configurations within a predefined neighbor graph). We estimate \(\sigma_{\rm num}\) and \(\sigma_{\rm mod}\) to quantify numerical noise and model differences, respectively. \(\sigma_{\rm num}\) is computed from the absolute differences between repeated runs of the same configuration, using the MAD and scaled by a factor of 1.48 which is to convert the MAD into an approximate standard deviation, providing a more accurate estimate of the noise in the data. \(\sigma_{\rm mod}\) is calculated from the differences between adjacent configurations' BIC values, also using MAD and scaling by 1.48. Both estimations help reduce sensitivity to small, non-physical fluctuations in BIC, ensuring robustness across different configurations and preserving the usual \(e^{-\Delta \mathrm{BIC}/2}\) behavior in the zero-noise limit. The combined noise scale \(\sigma\) is then:
\begin{equation}
\sigma \;=\; \sqrt{\sigma_{\rm mod}^2 \;+\; \sigma_{\rm num}^2 }.
\end{equation}
For each chain \(i\), with \(\Delta\mathrm{BIC}_i\), we define a latent variable \(X_i\), distributed as a normal variable:
\begin{equation}
X_i \;\sim\; \mathcal N\!\left( \mu_i,\, \sigma^2 \right), \qquad \mu_i \equiv -\tfrac{1}{2}\,\Delta\mathrm{BIC}_i.
\end{equation}
The effective (unnormalized) weight is the expectation:
\begin{equation}
\tilde w_i \;=\; \mathbb E\!\left[ \min\!\big(1, e^{X_i}\big) \right]
\;=\; \Phi\!\left(\frac{\mu_i}{\sigma}\right) \;+\; \exp\!\left(\mu_i + \tfrac{1}{2}\sigma^2\right)
\,\Phi\!\left(\frac{-\mu_i - \sigma^2}{\sigma}\right),
\end{equation}
where \(\Phi\) is the standard normal cumulative distribution function (CDF). The final weight is normalized by:
\begin{equation}
w_i \;=\; \frac{\tilde w_i}{\max_j \tilde w_j}.
\end{equation}
This weighting scheme is not intended to redefine Bayesian evidence or to replace a full evidence-based model comparison. Instead, it is designed to stabilize the relative ranking of closely related model configurations against numerical fluctuations arising from finite optimization accuracy and sampling noise, while preserving the usual \(\exp(-\Delta\mathrm{BIC}/2)\) behaviour in the zero-noise limit.

In practice, we compute the model-averaged parameter posterior distributions as:
\begin{equation}
P(\xi)_{\rm avg} = \sum_i w_i P(\xi | \mathcal{M}_i),
\end{equation}
where \(P(\xi | \mathcal{M}_i)\) is the posterior distribution of the parameter \(\xi\) for model \(\mathcal{M}_i\), and \(w_i\) is the weight for each model as determined by its BIC score. This method provides a model-averaged estimate for the parameters that accounts for the relative plausibility of each model. Table~\ref{tab:bic} presents the final BIC results for each chain in both the EPL and BPL families. 

\begin{table}
\centering
\caption{BIC comparison for EPL and BPL models. BIC values are not to be compared across model families.}
\label{tab:bic}
\begin{tabular}{ccccc}
\hline
$n_{\max}$ & Run & $\Delta$BIC & Weight \\
\hline
\multicolumn{4}{c}{EPL} \\
\hline
$\{9, 13\}$ & 2 & 0 & 1.00 \\
$\{9, 13\}$ & 1 & 39 & 0.94 \\
$\{7, 11\}$ & 2 & 541 & 0.30 \\
$\{7, 11\}$ & 1 & 542 & 0.30 \\
$\{8, 12\}$ & 2 & 640 & 0.22 \\
$\{8, 12\}$ & 1 & 647 & 0.21 \\
\hline
\multicolumn{4}{c}{BPL} \\
\hline
$\{9, 13\}$ & 1 & 0 & 1.00 \\
$\{9, 13\}$ & 2 & 2 & 1.00 \\
$\{7, 11\}$ & 2 & 639 & 0.22 \\
$\{7, 11\}$ & 1 & 650 & 0.21 \\
$\{8, 12\}$ & 2 & 998 & 0.05 \\
$\{8, 12\}$ & 1 & 1052 & 0.04 \\
\hline
\end{tabular}
\end{table}

\subsection{Combining Stellar Kinematics and External Convergence}
\label{comlos}

To constrain the mass-sheet degeneracy, we combine stellar kinematics with an external convergence (\(\kappa_{\text{ext}}\)) constraint. We first perform dynamical modeling of the deflector using Eq.~\ref{eq:losmodel}. The luminosity density \(l(r)\) is obtained by deprojecting the F160W surface brightness profile, which yields an effective radius \(\theta_{\text{eff}} = 3\farcs2 \pm 0\farcs02\). This profile is approximated with an elliptical multi-Gaussian expansion (MGE) \citep{Cappellari2002} and converted to a spherical 3D light profile under the Jeans equation assumption. A \(2\%\) uncertainty is propagated from \(\theta_{\text{eff}}\) to the MGE scales.

For the velocity anisotropy, we adopt a Jeffreys' prior on the scaling factor \(a_{\mathrm{ani}}\) (where \(r_{\mathrm{ani}} \equiv a_{\mathrm{ani}} \, \theta_{\mathrm{eff}}\)), sampling \(\log a_{\mathrm{ani}}\) uniformly over \([0.5, 5]\,\theta_{\mathrm{eff}}\) following \citet{Birrer2016,Birrer2020}.

The external convergence \(\kappa_{\mathrm{ext}}\) is estimated statistically from the observed galaxy environment following the TDCOSMO/STRIDES line-of-sight methodology. In practice, we extract from \citet{Shajib2022} the GLEE-based \(P(\kappa_{\mathrm{ext}}\mid\gamma_{\mathrm{ext}})\) distributions for WGD~2038--4008, which were constructed from the observed environment using the procedure of \citet{BuckleyGeer2020}. An important point is that the bins with larger \(\gamma_{\mathrm{ext}}\) in these distributions preferentially correspond to larger \(\kappa_{\mathrm{ext}}\), in several cases reaching \(\kappa_{\mathrm{ext}}\gtrsim 0.1\). Physically, such a large external field would generally require a fairly massive nearby group or cluster, rather than being typical of ordinary cosmic shear. Motivated by this concern, we further checked the redMaPPer catalogue in the vicinity of WGD~2038--4008. We summarize in Table~\ref{tab:redmapper_env} the three nearby systems identified in this check: two systems within \(\sim 3\) arcmin that were already noted by \citet{Shajib2022}, and the richest system found within \(10\) arcmin in our additional search. While this comparison does not exclude all line-of-sight contributions, it does not indicate the presence of an obvious very massive cluster immediately adjacent to the lens that would justify forcing the inference toward the highest-\(\gamma_{\mathrm{ext}}\) bins. We therefore do not weight the \(P(\kappa_{\mathrm{ext}}\mid\gamma_{\mathrm{ext}})\) distributions by the lens-model shear posterior. Instead, we assign equal weight to each \(\gamma_{\mathrm{ext}}\) bin and marginalize over them, thereby constructing an environment-only prior \(P(\kappa_{\mathrm{ext}})\) that is adopted identically for both the EPL and BPL model families.

{\color{blue}
\begin{table}
\centering
\caption{Nearby redMaPPer systems around WGD~2038--4008 (\(\{\mathrm{RA}, \mathrm{Dec}\} = \{309.511379, -40.137024\}\)) used as a qualitative environmental check. The first two entries are the systems reported by \citet{BuckleyGeer2020} within \(\sim 3\) arcmin of the lens, while the third entry is the richest redMaPPer system found within \(10\) arcmin in our additional search. The columns show the cluster ID, cluster redshift \(z\), richness \(\lambda\) and projected distance of the centroid to the lens \(\Delta\theta \) (arcmin).}
\label{tab:redmapper_env}
\begin{tabular}{cccc}
\hline
MEM\_MATCH\_ID & \(z\) & \(\lambda\) & \(\Delta\theta\) \\
\hline
62659  & \(0.221 \pm 0.008\) & \(5.1 \pm 1.7\)   & 1.798 \\
138669 & \(0.405 \pm 0.017\) & \(10.8 \pm 2.0\)  & 1.810 \\
27780  & \(0.238 \pm 0.006\) & \(22.8 \pm 3.2\)  & 4.870 \\
\hline
\end{tabular}
\end{table}
}

The distance ratio \(D_{\mathrm{s}}/D_{\mathrm{ds}}\) is constrained using the Pantheon SN sample \citep{Scolnic2018}. We approximate the luminosity distance with a fourth-order Taylor expansion, select the highest-evidence model including \(j_0\) via nested sampling \citep{Skilling2004}, and convert to angular diameter distances via \(D_A = D_L/(1+z)^2\) to obtain the \(D_{\mathrm{s}}/D_{\mathrm{ds}}\) posterior.

Following \citet{Birrer2016}, the aperture velocity dispersion under internal and external MST corrections can be written as
\begin{equation}
\sigma_{\mathrm{ap}}^2
=
(1-\kappa_{\mathrm{ext}})
\lambda_{\mathrm{int}}
\left(\frac{D_{\mathrm{s}}}{D_{\mathrm{ds}}}\right)
c^2
J(\xi_{\mathrm{lens}}, \xi_{\mathrm{light}}, \beta_{\mathrm{ani}}),
\end{equation}
where \(J(\xi_{\mathrm{lens}}, \xi_{\mathrm{light}}, \beta_{\mathrm{ani}})\) is a dimensionless factor that depends only on angular quantities, namely the lens-model parameters \(\xi_{\mathrm{lens}}\), the light distribution \(\xi_{\mathrm{light}}\), and the OM anisotropy \(\beta_{\mathrm{ani}}\), and is directly related to the model convergence \(\kappa_{\mathrm{model}}\) \citep{Birrer2016}. Rearranging this expression gives the internal mass normalization as
\begin{equation}
\lambda_{\mathrm{int}} =
\frac{\sigma_{\mathrm{ap}}^2}{(1 - \kappa_{\mathrm{ext}})\left( \frac{D_{\mathrm{s}}}{D_{\mathrm{ds}}} \right)c^2 J(\xi_{\mathrm{lens}}, \xi_{\mathrm{light}}, \beta_{\mathrm{ani}})}.
\end{equation}

To make the role of the LOS term explicit, we additionally consider the diagnostic case \(\lambda_{\mathrm{int}}=1\), that is, no internal mass-sheet correction. In this case, the stellar kinematics are used only to assess whether the LOS-corrected lens-model prediction is consistent with the observed \(\sigma_{\mathrm{ap}}\). Operationally, this is equivalent to importance-weighting the lens-model posterior with the kinematic likelihood computed for \(\lambda_{\mathrm{int}}=1\), using the same anisotropy prescription, the same environment-only \(P(\kappa_{\mathrm{ext}})\), and the same \(D_{\mathrm{s}}/D_{\mathrm{ds}}\) prior. In other words, the total correction factor is reduced to \((1-\kappa_{\mathrm{ext}})\) alone, and no separate internal MST freedom is inferred.



\begin{figure*}
    \centering
    \includegraphics[width=1.15\textwidth]{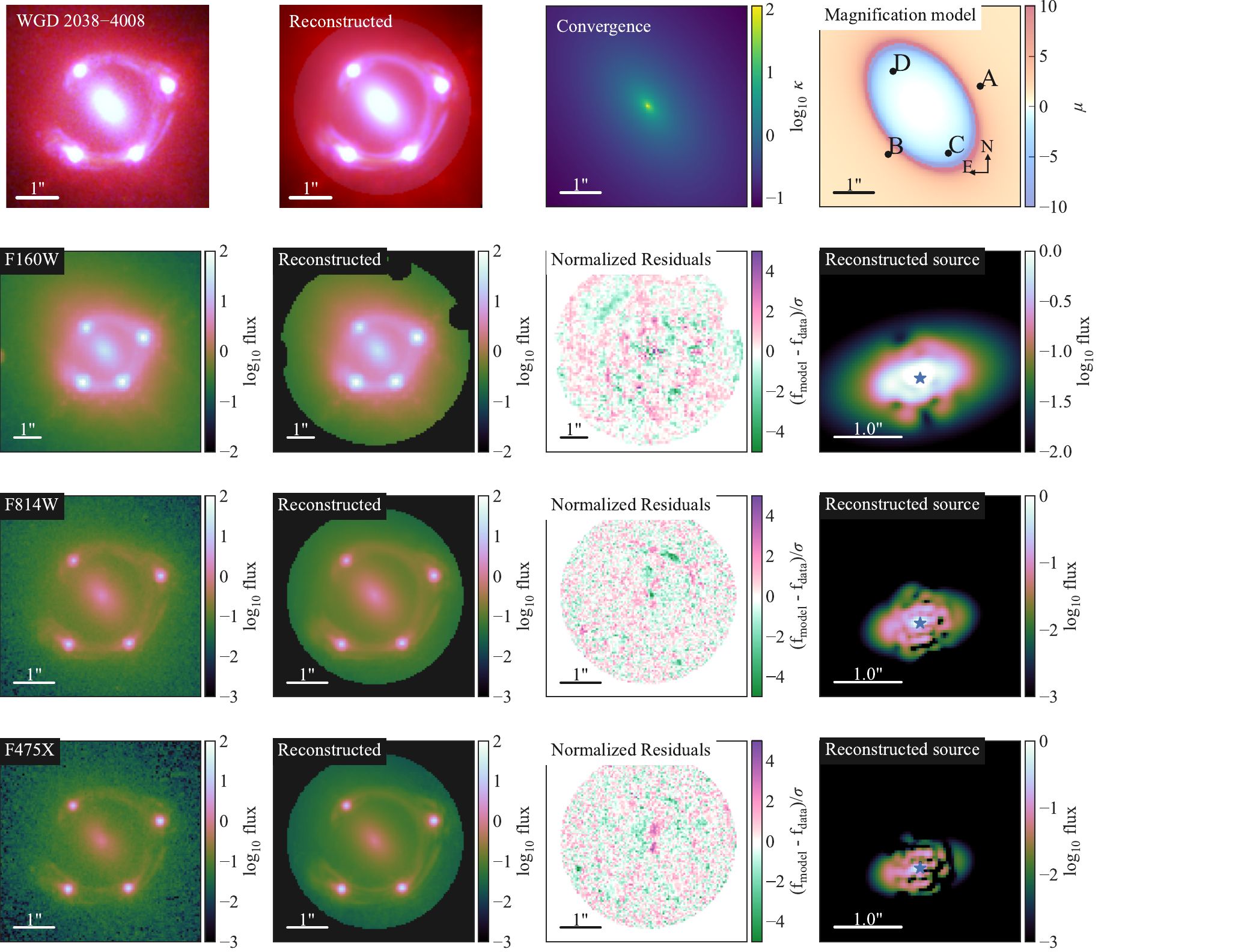}
    \caption{Lenstronomy-based lens model and image reconstruction of WGD 2038-4008 using an EPL mass profile. The top row shows, from left to right, the observed RGB composite, the model-predicted RGB composite, the convergence map \((\kappa)\), and the magnification \((\mu)\). Rows 2-4 display, for each HST filter, the observed image, the reconstructed image, the residual, and the reconstructed source plane: F160W (row 2), F814W (row 3), and F475X (row 4). All scale bars correspond to \(1^{\prime\prime}\). In the source panels, the star marks the centroid of the quasar host galaxy.}
    \label{fig:wgd_pemd_fit}
\end{figure*}

\begin{figure*}
    \centering
    \includegraphics[width=1.15\textwidth]{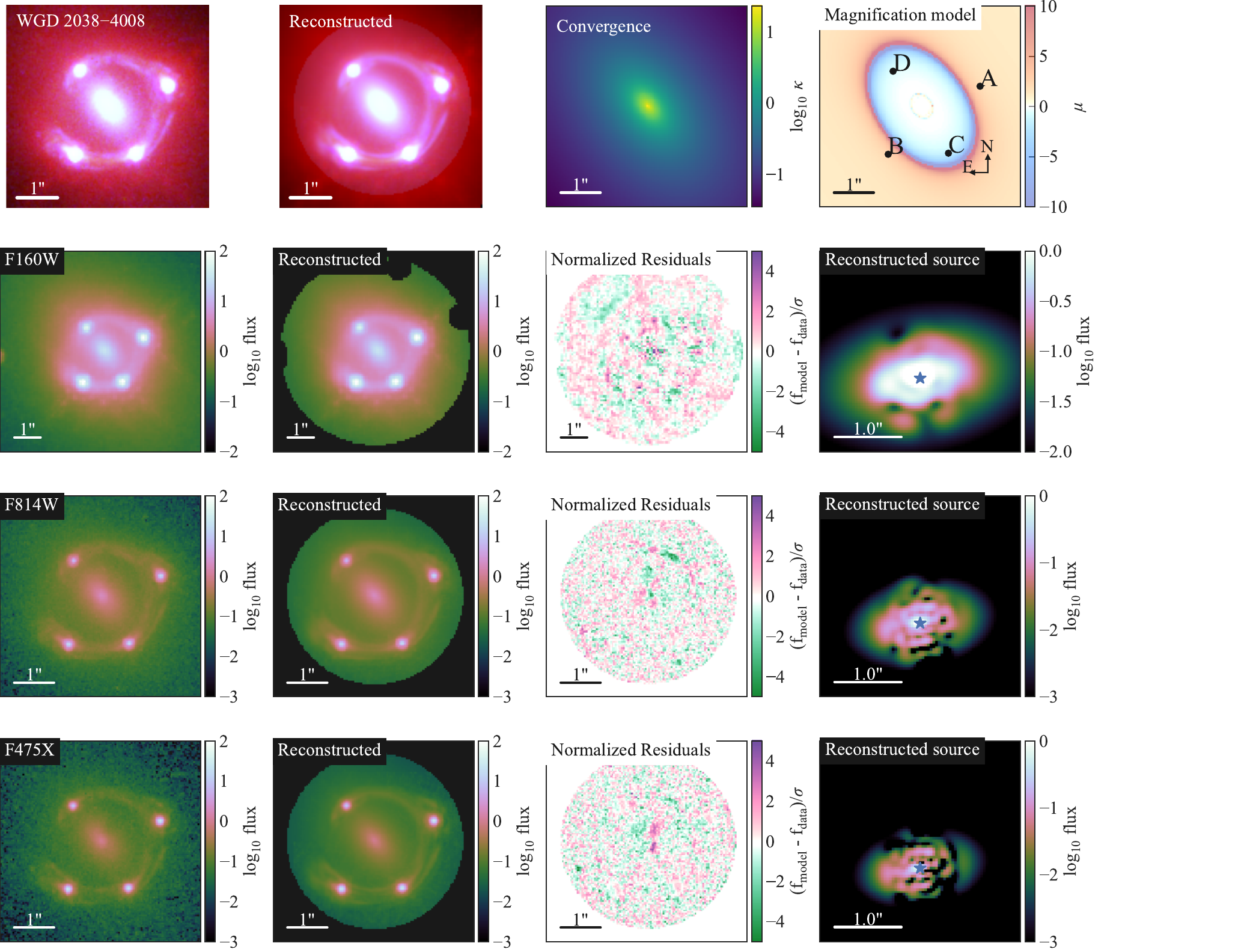}
    \caption{Same as Fig.~\ref{fig:wgd_pemd_fit}. Lenstronomy-based lens model and image reconstruction of WGD~2038--4008 using the BPL mass profile. The magnification map exhibits a nearly circular feature near the lens centre, indicating that the central lensing geometry is primarily governed by the mass components with lower slope. In the absence of a clearly detectable central image, the strong-lensing imaging observables mainly constrain the projected mass distribution near the Einstein radius, while providing only weak sensitivity to the mass profile at radii well inside it. As a result, the inferred inner mass slope is not directly constrained by the imaging data alone, but is largely driven by the assumed functional form of the mass parameterization.}
    \label{fig:wgd_bpl_fit}
\end{figure*}

\section{Results}
\label{sec:results}

In this section, we present the main outcomes of our analysis of WGD~2038--4008, focusing on the comparison between the EPL and BPL mass-model families. We begin by comparing the lens model fits and derived structural parameters for the two models, including their Einstein radii, projected ellipticities, external shear, and radial slope structure in Sect.~\ref{sec:fitting_results}. We then examine the predicted time delays based on image modeling alone under a fiducial flat $\Lambda$CDM cosmology in Sect.~\ref{sec:time_delay_image_only}. Following this, we incorporate stellar kinematic constraints and line-of-sight (LOS) external convergence to refine the time-delay predictions and obtain mass-sheet factors in Sect.~\ref{sec:time_delay_with_kinematics_los}. We next use the BIC-weighted posteriors from the combined imaging, kinematics, and LOS analyses to construct the time-delay distance and \(H_0\) posteriors, quantifying the model dependence of the cosmographic inference between the EPL and BPL families in Sect.~\ref{sec:h0_posteriors}. We also present a diagnostic case with fixed \(\lambda_{\rm int}=1\) in Appendix~\ref{app:lambda1}.

\subsection{Fitting Results and Parameter Constraints}
\label{sec:fitting_results}
In this subsection, we present the lens model fitting results for both the EPL and BPL model families using the multi-band imaging data. Figs.~\ref{fig:wgd_pemd_fit} and \ref{fig:wgd_bpl_fit} show the observed images, model predictions, residuals, and reconstructed source for the best-fitting models in each family. Overall, both models reproduce the observed lensed morphology and image configurations equally well. The residual patterns and reconstructed source structures are broadly similar between the two model families, and no statistically significant differences are evident at the current data quality and angular resolution. Consequently, the imaging data alone do not provide sufficient discriminatory power to distinguish between the EPL and BPL mass profiles.  Table~\ref{tab:chi2_epl_bpl_multiband} summarizes the reduced \(\chi^2\) values for the joint three-band and for each band separately (F160W, F814W, and F475X) for the best-fitting models shown in Figs.~\ref{fig:wgd_pemd_fit} and \ref{fig:wgd_bpl_fit}, indicating that the overall fit quality is statistically comparable between the two model families: the combined reduced \(\chi^2\) values differ only marginally and are similarly close to unity, while the band-by-band reduced \(\chi^2\) values differ at the \(10^{-3}\sim 10^{-2}\) level (including in F160W), with the optical bands (F814W and F475X) effectively indistinguishable.

\begin{table}
\centering
\caption{Reduced \(\chi^2\) for EPL and BPL best-fitting models (shown in Figs.~\ref{fig:wgd_pemd_fit} and \ref{fig:wgd_bpl_fit}) for all the three bands combined, and for the individual bands F160W, F814W, and F475X.}
\label{tab:chi2_epl_bpl_multiband}
\begin{tabular}{ccc}
\hline
Bands & BPL & EPL \\
\hline
Combined         & $0.907$ & $0.903$ \\
F160W            & $0.752$  & $0.738$ \\
F814W            & $0.900$  & $0.898$ \\
F475X            & $1.013$  & $1.014$ \\
\hline
\end{tabular}
\end{table}

\begin{figure*}
    \centering
    \includegraphics[width=0.95\textwidth]{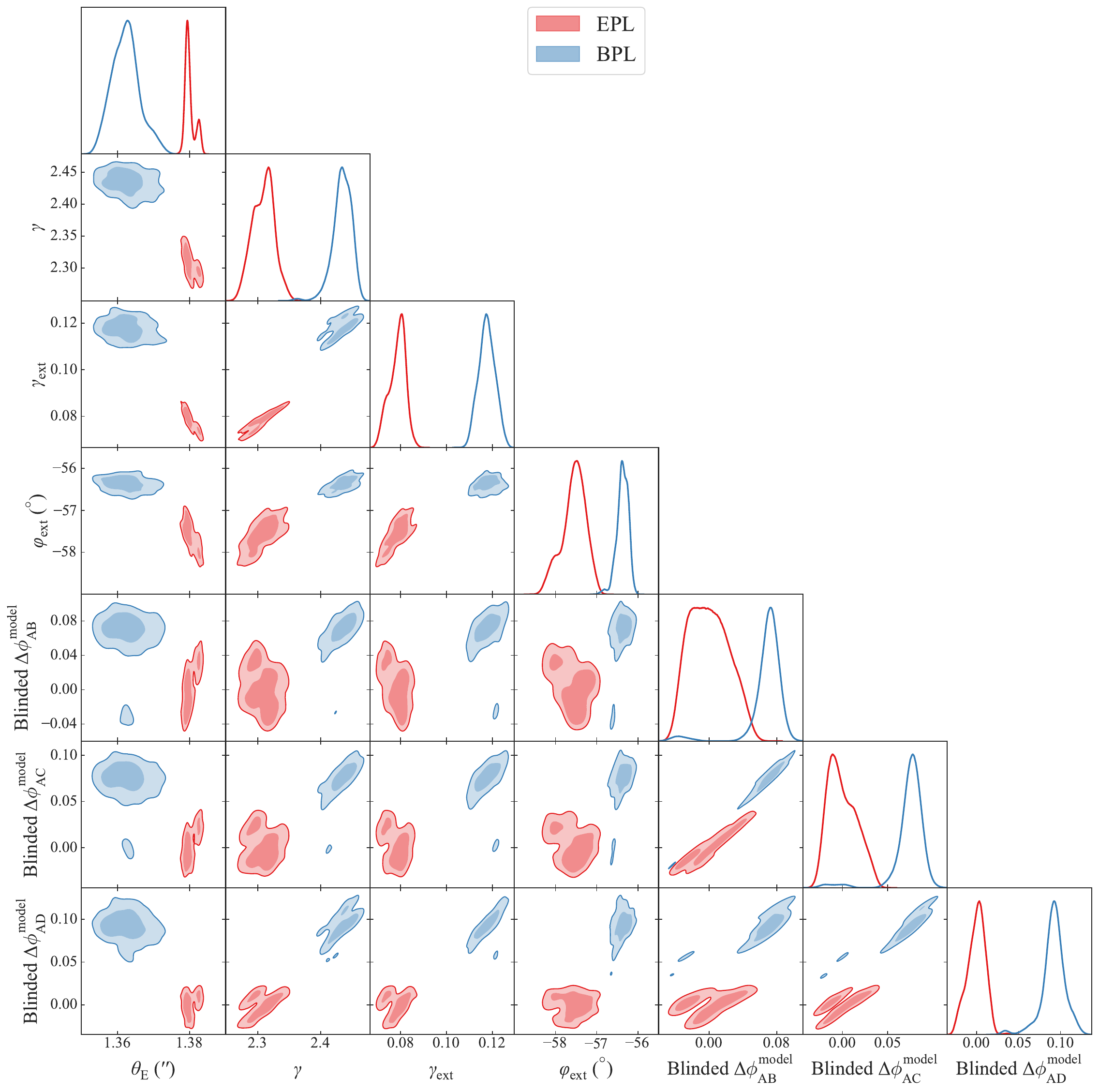}
    \caption{ distributions of BIC weighted parameters of the EPL (red) and BPL (blue) models, specifically focusing on the effective Einstein radius \(\theta_{\rm E}\). The tangential critical curve is traced at points where the tangential eigenvalue of the lensing Jacobian equals zero. Subsequently, the area of the primary closed loop encircling the lens center is determined. The local “effective” slope is computed with \(\gamma(\theta_{\rm E})\;=\;2\;-\;\frac{{\rm d}\ln \alpha(\theta_{\rm E})}{{\rm d} r}\,\)where \(\alpha(\theta_{\rm E})\) is the azimuthally averaged deflection amplitude; thus \(\gamma(\theta_{\rm E})\) is a lensing-derived (2D) effective slope obtained from the deflection field (for a true power-law, it equals the usual logarithmic index). To compare the differences in Fermat-potential offsets among the image positions, we set the mean value of the EPL model’s \(\phi_{A X}\) as the zero point of the coordinate reference. Shaded bands indicate the \(1\sigma\) and \(2\sigma\) credible intervals. } 
    \label{fig:wgd_maps}
\end{figure*}

Fig.~\ref{fig:wgd_maps} and Table~\ref{tab:wgd2038_posteriors} present the BIC–weighted posteriors for key lensing diagnostics and compare the projected mass distribution (convergence \( \kappa \)) and magnification \( \mu \) of the EPL and BPL families. The effective Einstein radius is \( \theta_{\rm E} = 1\farcs380 \pm 0\farcs001 \) for EPL and \( \theta_{\rm E} = 1\farcs364 \pm 0\farcs004 \) for BPL. The mass ellipticity at intermediate radii is similar, with \( q_{\rm m} = 0.600 \pm 0.005 \) (EPL) and \( 0.650 \pm 0.02 \) (BPL), i.e., \( 1 - q_{\rm m} \approx 0.35 \sim 0.4\). The inferred external shear is strongly model dependent: we find \(\gamma_{\rm ext} = 0.077 \pm 0.004\) for the EPL family, compared to \(0.119 \pm 0.004\) for the BPL family, implying a \(\simeq 7.4\sigma\) discrepancy that far exceeds the nominal statistical uncertainties. Given the well-known degeneracies between \(\gamma_{\rm ext}\) and other lens parameters, and its sensitivity to modelling choices (most notably PSF reconstruction), we interpret this offset as reflecting model-dependent trade-offs rather than a purely environmental measurement \citep{Shajib2022}.

The main distinction concerns the radial slope structure. The single-slope EPL corresponds to a density index \( \gamma' = 2.30 \pm 0.02 \), whereas the BPL model is consistent with a broken profile characterized by an outer slope \( \alpha = 2.799 \pm 0.066 \), an inner slope \( \alpha_c = 1.26 \pm 0.06 \), and a break radius \( r_c = 0.40 \pm 0.08 \, \mathrm{arcsec} \). As a result, the BPL parameterization allows for a shallower inner mass distribution within \( \sim 1.2\!-\!1.8 \,\mathrm{kpc} \) (this is not directly constrained by imaging), while remaining steeper at larger radii. Consistent with this behaviour, the lensing-derived effective slope evaluated at the Einstein ring, \( \gamma(\theta_{\rm E}) \), is modestly steeper for the BPL model (\( 2.43 \pm 0.02 \)) than for the EPL model (\( 2.30 \pm 0.02 \)), reflecting that \( \theta_{\rm E} \) lies closer to the outer-slope regime.

This behaviour is further illustrated by the radial profiles of the mean convergence shown in Sect.~\ref{app:psidv}. While the EPL and BPL models are tightly constrained to yield similar enclosed convergence near the Einstein radius—where strong-lensing imaging provides the strongest leverage—they exhibit different trends at smaller radii. In particular, the BPL model permits a redistribution of mass such that the inner regions (\( R \ll \theta_{\rm E} \)) are comparatively flatter, whereas the effective slope evaluated near \( \theta_{\rm E} \) can be steeper. This difference arises naturally from the radial flexibility of the BPL parameterization, rather than representing a contradiction between constraints on inner and Einstein-scale mass distributions.

While the inferred BPL inner slope is qualitatively compatible with a core-like flattening, we emphasize that, in the absence of a clearly detected central image in the current data, strong-lensing observables are primarily sensitive to the projected mass distribution near the Einstein radius and thus provide only weak constraints on the inner profile at radii \( r \ll \theta_{\rm E} \). In this regime, the inference is driven by the adopted mass parameterization and the joint lensing–dynamical modelling, rather than by lensing information alone. Overall, both model families provide comparably good fits to the data, with only subtle trade-offs in source morphology and inner-slope preference that the current observations do not decisively resolve.

Notably, the parameter results for the EPL model show some differences when compared to those obtained by \citep{Shajib2022} using \textsc{Lenstronomy}, though they are more consistent with the results derived from the \textsc{GLEE} code. These differences could stem from the inherent degeneracy of the EPL model and the random nature of PSF iteration. We extended the PSF iteration process two to three times longer than in previous studies, improving the overall accuracy of the model.

\subsection{Time-delay predictions in the image-modelling-only case}
\label{sec:time_delay_image_only}
To assess the model dependence of the predicted time delays prior to including kinematics and external convergence, we compute the BIC-weighted posteriors of the EPL and BPL families under a flat \(\Lambda\)CDM cosmology with \(H_0 = 70\,\rm km\,s^{-1}\,Mpc^{-1}\) and \(\Omega_m = 0.3\). 

The BIC-weighted time-delay posteriors for the AB, AC, and AD image pairs show that the two mass-model families yield systematically different predictions: the BPL model produces larger \(|\Delta t|\) than the EPL model across all image pairs. This behaviour reflects the different inner-density slopes of the two families, which reshape the radial mass profile and therefore modify the Fermat-potential surfaces in a coherent manner. In support of this argument, we provide a more detailed discussion based on the model posteriors in Sect.~\ref{app:psidv}. We find that the dominant contribution to the time-delay difference between the EPL and BPL families arises from the geometric term. The deflection-angle variations modify the geometric path length component of the Fermat potential in a coherent manner across all image pairs, while the contribution from differences in the lens potential remains subdominant.

Quantitatively, the raw EPL and BPL predictions (i.e.\ without the stellar kinematics and LOS correction) are:
\begin{equation}
\Delta t_{\rm AB}^{\rm EPL} = -5.543^{+0.139}_{-0.133}\,\rm days, \quad
\Delta t_{\rm AB}^{\rm BPL} = -5.945^{+0.080}_{-0.063}\,\rm days,
\end{equation}
\begin{equation}
\Delta t_{\rm AC}^{\rm EPL} = -11.198^{+0.165}_{-0.196}\,\rm days, \quad
\Delta t_{\rm AC}^{\rm BPL} = -12.085^{+0.149}_{-0.111}\,\rm days,
\end{equation}
\begin{equation}
\Delta t_{\rm AD}^{\rm EPL} = -27.314^{+0.293}_{-0.265}\,\rm days, \quad
\Delta t_{\rm AD}^{\rm BPL} = -29.866^{+0.356}_{-0.336}\,\rm days.
\end{equation}

The relative time delays differ from those of the EPL model by approximately \(9-16\%\). The offsets between the two families are therefore coherent, stable across image pairs, and directly trace the differing radial mass distributions. Since larger \(|\Delta t|\) at fixed observed delays corresponds to a smaller inferred time-delay distance \(D_{\Delta t}\), and thus a higher inferred \(H_0\), these discrepancies demonstrate that the choice between EPL and BPL already induces a non-negligible systematic shift at the image-modelling-only stage. This motivates treating the inner-profile uncertainty as an explicit component of the error budget in the full cosmographic analysis.

\subsection{Time-delay predictions combining kinematics and LOS}
\label{sec:time_delay_with_kinematics_los}

\begin{figure}
    \centering
    \includegraphics[width=\columnwidth]{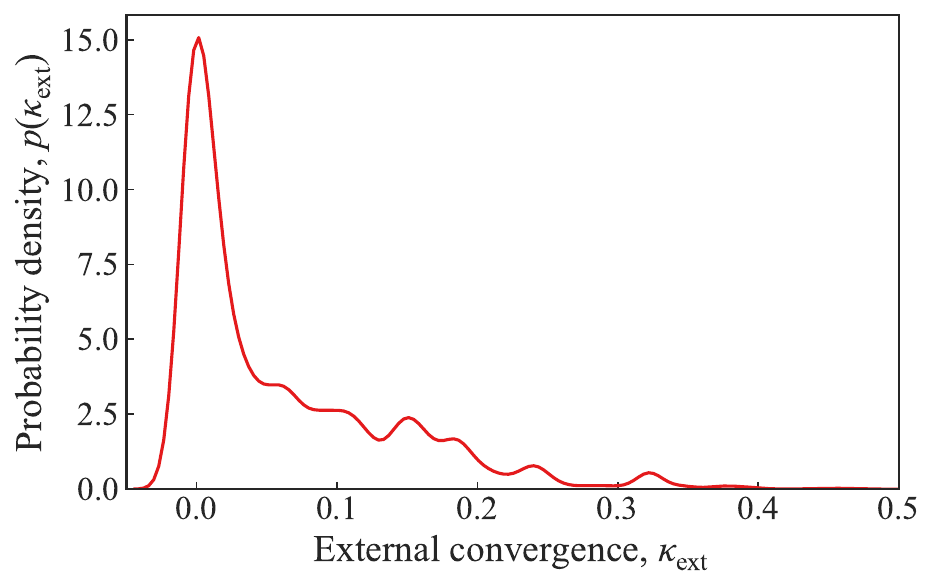}
    \caption{Environment-only external-convergence prior \(P(\kappa_{\rm ext})\) for WGD~2038--4008. This distribution is obtained by averaging the GLEE-based \(P(\kappa_{\rm ext}\mid\gamma_{\rm ext})\) curves of \citet{Shajib2022} with equal weight over external-shear bins. The same prior is adopted for both the EPL and BPL model families.}
    \label{fig:kappaext}
\end{figure}

We refine the lens-model inferences by incorporating both the stellar kinematics and the LOS external convergence. The \(\kappa_{\rm ext}\) prior is constructed from the observed line-of-sight environment following the procedure described in Sect.~\ref{comlos}. The same environment-only distribution \(P(\kappa_{\rm ext})\) is adopted for both the EPL and BPL model families (Fig.~\ref{fig:kappaext}).

\begin{figure}
  \centering
  \includegraphics[width=\columnwidth]{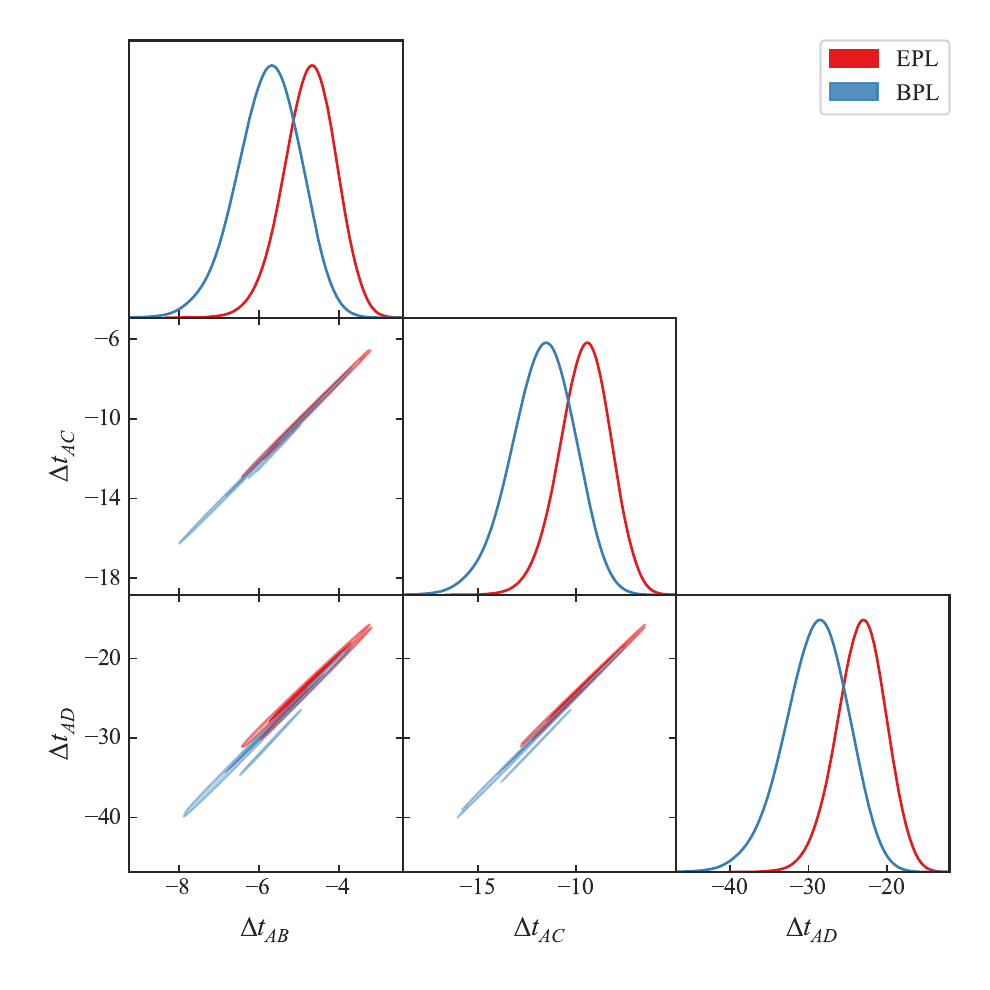}
  \caption{Corner plot of the model-predicted relative time delays (days) for the three pairs (AB, AC, AD). Chains are truncated to their last \(N=1000\) samples, run weights are distributed uniformly over retained samples, and BPL (red) and EPL (blue) families are overplotted with filled contours. (Assuming flat \(\Lambda\)CDM cosmology with \(H_0 = 70\,\rm km\,s^{-1}\,Mpc^{-1}\) and \(\Omega_m = 0.3\))}
  \label{fig:timedelay_corner}
\end{figure}

After applying the full mass-sheet correction \((1-\kappa_{\rm ext})\lambda_{\rm int}\) and combining all constraints, the predicted time delays exhibit coherent and non-negligible shifts relative to the image-modelling-only case. The corrected BIC-weighted posteriors are (Fig~\ref{fig:timedelay_corner})
\begin{equation}
\Delta t_{\rm AB}^{\rm EPL} = -4.713^{+0.627}_{-0.675}\,\rm days,\quad
\Delta t_{\rm AB}^{\rm BPL} = -5.734^{+0.775}_{-0.825}\,\rm days,
\end{equation}
\begin{equation}
\Delta t_{\rm AC}^{\rm EPL} = -9.527^{+1.250}_{-1.347}\,\rm days,\quad
\Delta t_{\rm AC}^{\rm BPL} = -11.657^{+1.573}_{-1.685}\,\rm days,
\end{equation}
\begin{equation}
\Delta t_{\rm AD}^{\rm EPL} = -23.216^{+3.039}_{-3.240}\,\rm days,\quad
\Delta t_{\rm AD}^{\rm BPL} = -28.862^{+3.921}_{-4.251}\,\rm days.
\end{equation}

\begin{figure}
\centering
\includegraphics[width=\columnwidth]{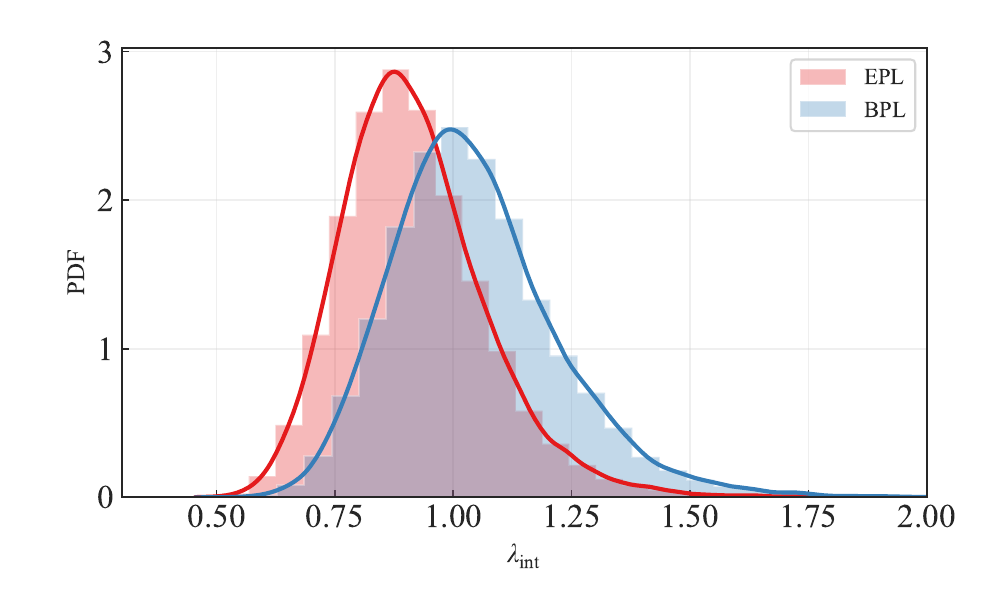}
\caption{Posterior of the mass-sheet factor \(\lambda_{\rm int}\) for the EPL and BPL model families. The EPL family predicts a median of \(0.901\), whereas the BPL family favours \(1.026\).}
\label{fig:kappalambda}
\end{figure}

It can be seen that the differences in the relative time delays become even more pronounced after incorporating the kinematics and LOS corrections, reaching up to about \(20-24\%\). To make explicit the effect of jointly sampling the internal mass profile and the LOS convergence, we compute the posterior distribution of the mass-sheet factor \(\lambda_{\rm int}\) (Fig.~\ref{fig:kappalambda}). The two families differ: 
the EPL model yields a median of \(0.901 \pm 0.155\), while the BPL model gives \(1.026 \pm 0.182\).  
These offsets demonstrate that the kinematics--LOS combination reshapes the effective time-delay scaling in a model-dependent manner, amplifying the differences already present in the image-modelling-only predictions.

Taken together, the corrected \(\Delta t\) posteriors and the inferred mass-sheet factors show that the two model families retain distinct cosmographic implications for the inferred time-delay distance and \(H_0\) even after incorporating all available non-imaging information. Although the stellar kinematics and the \(D_{\rm s}/D_{\rm ds}\) prior provide independent observational constraints, both \(\lambda_{\rm int}\) and \(\kappa_{\rm ext}\) are inferred quantities that remain conditional on the adopted lens mass model. Consequently, the resulting mass-sheet factor \((1-\kappa_{\rm ext})\lambda_{\rm int}\) should be interpreted as the outcome of a joint, model-dependent inference, rather than as an independent criterion for assessing the physical preference of a given mass-profile family.

\begin{table}
    \centering
    \caption{Comparison of BIC-weighted model parameters. Quoted values are medians and 68\% credible intervals. Note that here, \(\theta_E\) refers to the spherical-equivalent Einstein radius (i.e., effective Einstein radius). }
    \label{tab:wgd2038_posteriors}
    \begin{tabular}{lcc}
    \hline
    Parameter & EPL model & BPL model \\
    \hline
    $\gamma$ (slope) & $2.30 \pm 0.02$ & $2.43 \pm 0.02$ \\
    $\alpha$ (outer slope) & --- & $2.799 \pm 0.066$ \\
    $\alpha_c$ (inner slope) & --- & $1.26 \pm 0.06$ \\
    $r_c$ (arcsec) & --- & $0.40 \pm 0.08$ \\
    $\theta_{\rm E}$ (arcsec) & $1.380 \pm 0.001$ & $1.364 \pm 0.004$ \\
    $q_{\rm m}$ (axis ratio) & $0.600 \pm 0.005$ & $0.650 \pm 0.02$ \\
    $\gamma_{\rm ext}$ & $0.077 \pm 0.004$ & $0.119 \pm 0.004$ \\
    \hline
    \end{tabular}
\end{table}

\subsection{Combine time-delay measurements}
\label{sec:h0_posteriors}
We now construct the time-delay distance posterior by combining the BIC--weighted posterior samples of the lens-model parameters and external convergence with the observed time delays \citep{Wong2024}. In that work, the time delays and their full covariance were inferred with the \textsc{PyCS3} \citep{Millon2020pycs} curve-shifting analysis, where repeated optimizations with randomized initial shifts and mock light-curve realizations (including microlensing-like variability) yield a joint posterior from which the covariance is estimated; we therefore adopt the covariance matrix for delays relative to image A, which contains the full time-delay information. The time delays between images AB, AC, and AD are $-12.4$, $-5.3$, and $-33.3$ days, and the covariance matrix $\mathbf{C}$ is shown in Table~\ref{tab:delay_covariance_matrix}.

\begin{table}
\centering
\caption{Covariance matrix $\mathbf{C}$ for the time delays between quasar images A, B, C, and D (units: days$^2$).}
\label{tab:delay_covariance_matrix}
\begin{tabular}{cccc}
\hline
\textbf{Covariance (days\(^2\))} & \textbf{AB} & \textbf{AC} & \textbf{AD} \\
\hline
\textbf{AB} & 14.2 & 6.1 & 7.5 \\
\hline
\textbf{AC} & 6.1 & 14.8 & 7.1 \\
\hline
\textbf{AD} & 7.5 & 7.1 & 39.9 \\
\hline
\end{tabular}
\end{table}

\begin{figure}
  \centering
  \includegraphics[width=\columnwidth]{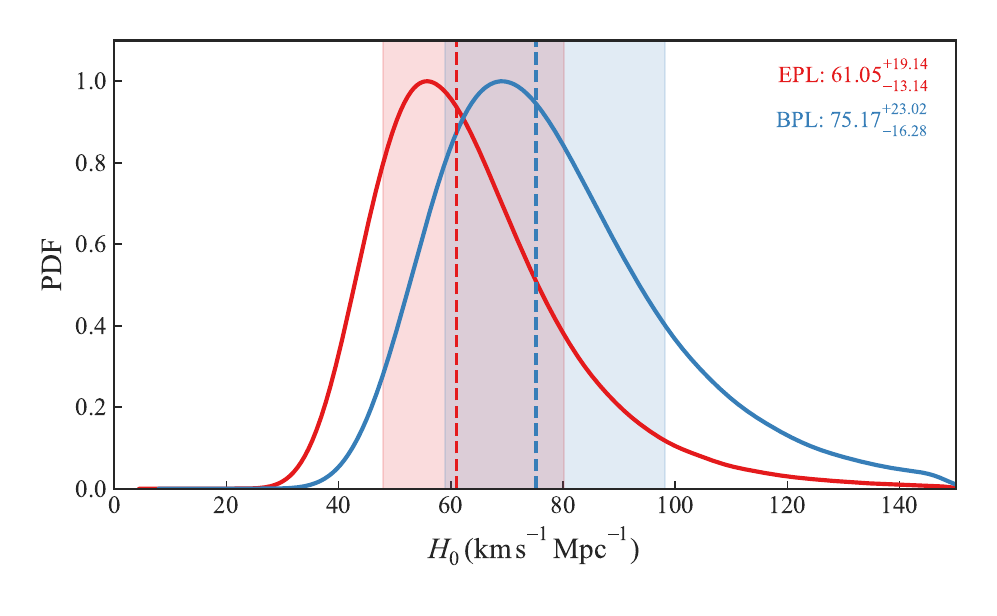}
    \caption{Posteriors of \(H_0\) obtained by mapping the sampled \(D_{\Delta t}\) distribution to the Hubble constant under a flat \(\Lambda\)CDM cosmology with fixed \(\Omega_{\rm m}=0.3\). Because \(D_{\Delta t}\propto H_0^{-1}\) at fixed \((z_{\ell},z_s)=(0.228,0.777)\), the \(H_0\) posterior follows directly from a change of variables applied to the sampled \(D_{\Delta t}\) distribution. The inferred differences between the EPL (red) and BPL (blue) families reflect the model-dependent impact of their respective \((1-\kappa_{\rm ext})\lambda_{\rm int}\) corrections. Medians and \(68\%\) credible intervals are annotated.
    }
  \label{fig:h0_family}
\end{figure}

For each lens model sampling points we have the predicted (and MST–corrected) time delays \(\Delta\mathbf{t}_{\rm fid}\) under a fixed fiducial cosmology. In a general cosmology, the theoretical delays scale linearly with the time-delay distance,
\begin{equation}
\Delta\mathbf{t}_{\rm th}(D_{\Delta t})
= R\,\Delta\mathbf{t}_{\rm fid}, \qquad
R \equiv \frac{D_{\Delta t}}{D_{\Delta t}^{\rm fid}},
\end{equation}
where \(D_{\Delta t}^{\rm fid}\) is computed under the fiducial flat
\(\Lambda\)CDM cosmology as
\begin{equation}
D_{\Delta t}^{\rm fid}
= \left[(1+z_{\rm d})\,
\frac{D_{\rm d}D_{\rm s}}{D_{\rm ds}}\right]_{\rm fid}.
\end{equation}
The likelihood of the observed time-delay vector 
\(\mathbf{d}\equiv\Delta\mathbf{t}_{\rm obs}\), with covariance matrix \(\mathbf{C}\), can be written as
\begin{equation}
\mathcal{L}_{\Delta\mathbf{t}_{\rm obs}}\!\left(
\mathcal{D}_{\Delta\mathbf{t}_{\rm obs}}\mid
\boldsymbol{\xi}_{{\rm model}},
\kappa_{{\rm ext}},
D_{\Delta t}
\right)
\propto
\exp\!\left[-\tfrac12\,\chi^2(R)\right],
\end{equation}
where \(R \equiv D_{\Delta t}/D_{\Delta t}^{\rm fid}\) and
\begin{equation}
\chi^2(R) =
\bigl(\mathbf{d}-R\,\Delta\mathbf{t}_{\rm fid}\bigr)^{\!T}
\mathbf{C}^{-1}
\bigl(\mathbf{d}-R\,\Delta\mathbf{t}_{\rm fid}\bigr).
\end{equation}
The full posterior of the time-delay distance is obtained by marginalising
over the lens-model and LOS degrees of freedom:
\begin{equation}
\begin{aligned}
p(D_{\Delta t}\mid\mathcal{D}) \propto 
\int 
&\mathcal{L}_{\Delta\mathbf{t}_{\rm obs}}\!\left(
\mathcal{D}_{\Delta\mathbf{t}_{\rm obs}}\mid 
\boldsymbol{\xi}_{\rm model},
\kappa_{\rm ext},
D_{\Delta t}
\right)\,
\\[2pt]
&\times\;
p\!\left(
\boldsymbol{\xi}_{\rm model},
\kappa_{\rm ext}\mid
\mathcal{D}_{\rm img,kin,LOS}
\right)\,
\mathrm{d}\boldsymbol{\xi}_{\rm model}\,
\mathrm{d}\kappa_{\rm ext}\;,
\end{aligned}
\label{eq:marg_Ddt}
\end{equation}
where the distribution  
\(p(\boldsymbol{\xi}_{\rm model},\kappa_{\rm ext}\mid
\mathcal{D}_{\rm img,kin,LOS})\)  
is represented numerically by the BIC–weighted posterior samples from the
imaging, kinematic, and LOS inference. In practice, this marginalisation
is performed by importance sampling over these model samples and summing
their contributions:
\begin{equation}
p(D_{\Delta t}\mid\mathcal{D})
\propto
\sum_{n} 
w_n\,
\mathcal{L}_{\Delta\mathbf{t}_{\rm obs}}
\bigl(\mathcal{D}_{\Delta\mathbf{t}_{\rm obs}}\mid 
\boldsymbol{\xi}_{{\rm model},n},
\kappa_{{\rm ext},n},
D_{\Delta t}
\bigr),
\end{equation}
with \(w_n\) denoting the BIC weights.
Since in our analysis the cosmology is restricted to flat \(\Lambda\)CDM with fixed \(\Omega_{\rm m}\), the mapping between \(D_{\Delta t}\) and \(H_0\) is strictly monotonic, \(D_{\Delta t}\propto H_0^{-1}\). The posterior of \(H_0\) therefore follows directly from a change of variables applied to the sampled \(D_{\Delta t}\) distribution. The resulting \(H_0\) posteriors are shown in Fig.~\ref{fig:h0_family}. Consistent with the discussion in Sect.~\ref{sec:time_delay_with_kinematics_los}, these differences reflect the model-dependent outcome of the joint lensing, kinematic, and LOS inference, rather than an independent validation of a specific mass-profile family. We emphasize that the observed shift in the inferred \(H_0\) between the EPL and BPL model families should be interpreted primarily as a manifestation of increased model flexibility under the mass-sheet degeneracy, rather than as direct evidence for a physical core in the inner mass distribution of the lens galaxy. While the BPL parameterization allows for a shallower inner slope, the current imaging and kinematic data do not uniquely constrain the mass profile at radii well below the effective Einstein radius. Consequently, the difference in \(H_0\) reflects how different model families explore the allowed degeneracy space when combined with stellar kinematics and external convergence constraints, rather than a definitive detection of core-like structure.

\section{Result Explanation}
\label{app:psidv}

As discussed in Sect.~\ref{sec:time_delay_image_only}, although the time delay is formally determined by differences in the Fermat potential, the dominant contribution to the time-delay difference between the EPL and BPL model families arises from the geometric term. In this section, we provide supplementary diagnostics of the lensing potential, deflection angles, and stellar-kinematic predictions to illustrate this point and to provide physical intuition for the model-dependent trends discussed in the main text.

\begin{figure}
\centering
\includegraphics[width=\columnwidth]{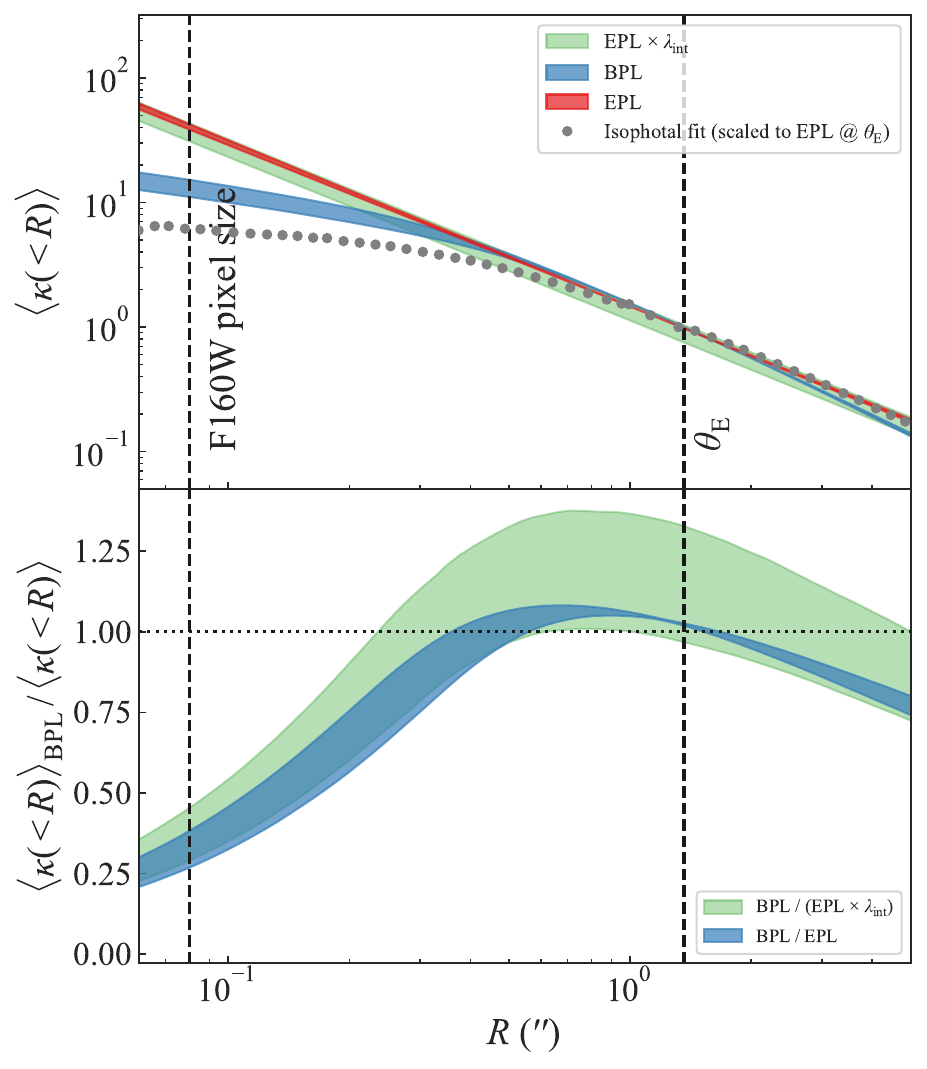}
\caption{Comparison of the azimuthally averaged convergence profile, \(\langle\kappa\rangle\), between the BPL and EPL models. The \textbf{top panel} shows the radial dependence of the mean projected mass within radius \(R\). The red band represents the EPL posterior, while the blue shaded region shows the BPL posterior. The green shaded region shows the EPL profile after rescaling by the sampled \(\lambda_{\rm int}\) distribution, i.e. EPL\(\times\lambda_{\rm int}\). The grey dots indicate the best-fitting isophotal light profile scaled to match the EPL normalization at the Einstein radius \(\theta_{\rm E}\). Vertical dashed lines denote the HST/F160W pixel scale (left) and the median Einstein radius (right). At small radii, the two models exhibit different inner slopes, while they converge near and beyond \(\theta_{\rm E}\). The rescaled EPL profile lies systematically below the original EPL profile and shows a trend that more closely resembles the BPL profile with $\lambda_{\rm int}=1$.
The \textbf{bottom panel} displays the ratios \(\langle\kappa(<R)\rangle_{\mathrm{BPL}} / \langle\kappa(<R)\rangle_{\mathrm{EPL}}\) (blue) and \(\langle\kappa(<R)\rangle_{\mathrm{BPL}} / \langle\kappa(<R)\rangle_{\mathrm{EPL}\times\lambda_{\rm int}}\) (green), highlighting the relative deviation between the two models as a function of radius. The shaded region encloses the \(1\sigma\) confidence interval for each corresponding posterior ratio.}
\label{fig:kappa_mean_comparison}
\end{figure}

Fig.~\ref{fig:kappa_mean_comparison} compares the azimuthally averaged projected mass distributions for the two model families. By construction, both models are constrained to reproduce the observed image configuration and the projected mass distribution near the Einstein radius. At smaller radii, however, the EPL and BPL models exhibit different inner radial behaviours, reflecting their distinct parameterizations of the mass profile. These differences lead to modest variations in the enclosed mass in the central regions, while the profiles converge near and beyond the Einstein radius. In addition to the original EPL and BPL profiles, Fig.~\ref{fig:kappa_mean_comparison} also shows the EPL profile after applying the internally inferred \(\lambda_{\rm int}\) rescaling. Once this internal correction is taken into account, the effective EPL profile in the lensing-sensitive region becomes more similar to the BPL profile than the uncorrected EPL profile. And in the lower panel, where the ratio relative to EPL\(\times\lambda_{\rm int}\) is correspondingly closer to unity than the ratio relative to the original EPL profile.

\begin{figure}
    \centering
    \includegraphics[width=\columnwidth]{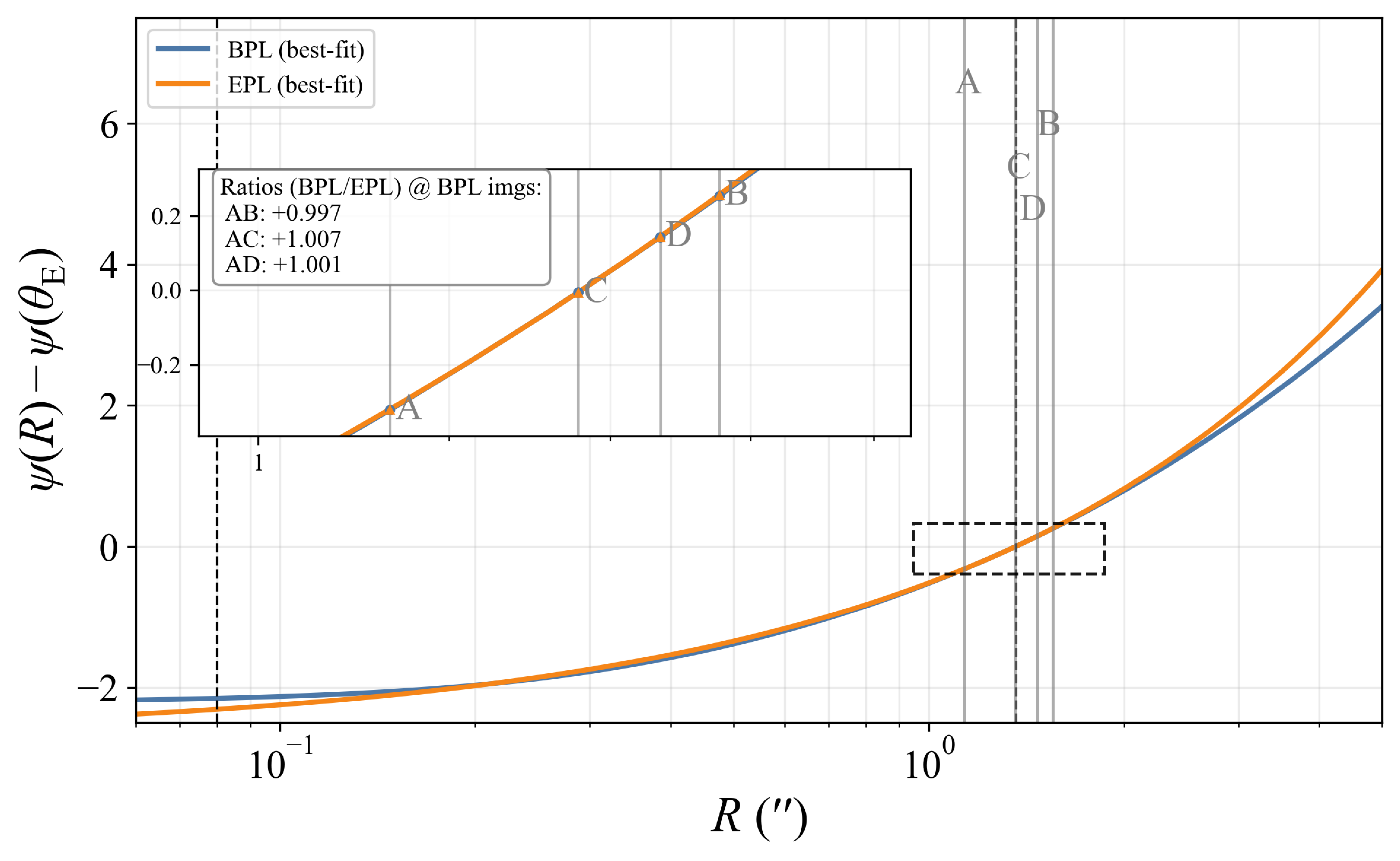}
    \caption{Lens potential comparison for image pairs AB, AC, and AD under the BPL and EPL models, using the best-fit parameters from the minimum-BIC realizations. The solid lines show the azimuthally averaged lens potential \(\psi(R)\), plotted relative to \(\psi(\theta_{\rm E})\), where \(\theta_{\rm E}\) is the median Einstein radius from the BPL BIC-weighted results. Vertical dashed lines mark the HST/F160W pixel scale (left) and \(\theta_{\rm E}\) (right), while the solid vertical lines indicate the image radii \(R_i\) (\(i=A,B,C,D\)). The inset highlights the region around the image positions.}
    \label{fig:lens_potential_comparison}
\end{figure}

To examine how these structural differences manifest at the image positions, we compute the lens potential and deflection angle profiles for both models using the best-fit parameters from the minimum-BIC realizations. The inferred image positions differ by less than \(0\farcs003\) between the two models, confirming that the image geometry is nearly identical. Consistent with this, the differences in the lens potential \(\Delta\psi\) evaluated at the image positions are very similar for the EPL and BPL models, with BPL-to-EPL ratios close to unity (see Fig.~\ref{fig:lens_potential_comparison}).

\begin{figure}
    \centering
    \includegraphics[width=\columnwidth]{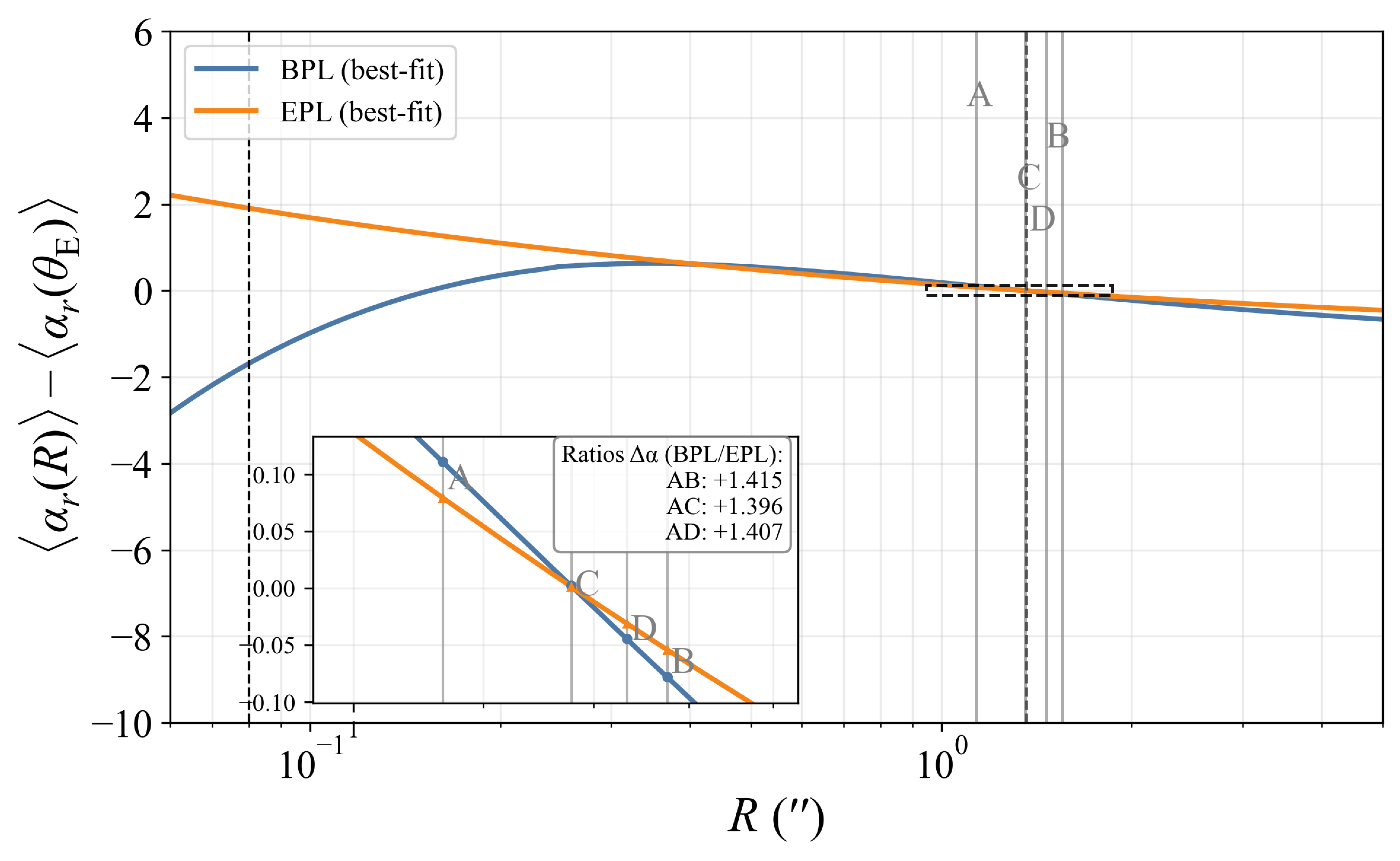}
    \caption{Deflection angle comparison for image pairs AB, AC, and AD under the BPL and EPL models, using the best-fit parameters from the minimum-BIC realizations. The solid lines show the azimuthally averaged deflection angles \(\alpha(R)\), plotted relative to \(\alpha(\theta_{\rm E})\). Vertical dashed lines mark the HST/F160W pixel scale (left) and \(\theta_{\rm E}\) (right), while the solid vertical lines indicate the image radii. The inset highlights the region around the image positions.}
    \label{fig:deflection_angle_comparison}
\end{figure}

In contrast, the deflection angle profiles exhibit more pronounced differences between the two model families at the image radii, as illustrated in Fig.~\ref{fig:deflection_angle_comparison}. These differences directly affect the geometric term of the Fermat potential and provide a natural explanation for why relatively small variations in the radial mass profile can translate into non-negligible differences in the model-predicted time delays, even when the lens potential itself remains similar at the image positions. These diagnostics therefore support the interpretation adopted in the main text that the time-delay differences are primarily driven by deflection-angle variations.

\begin{figure}
\centering
\includegraphics[width=\columnwidth]{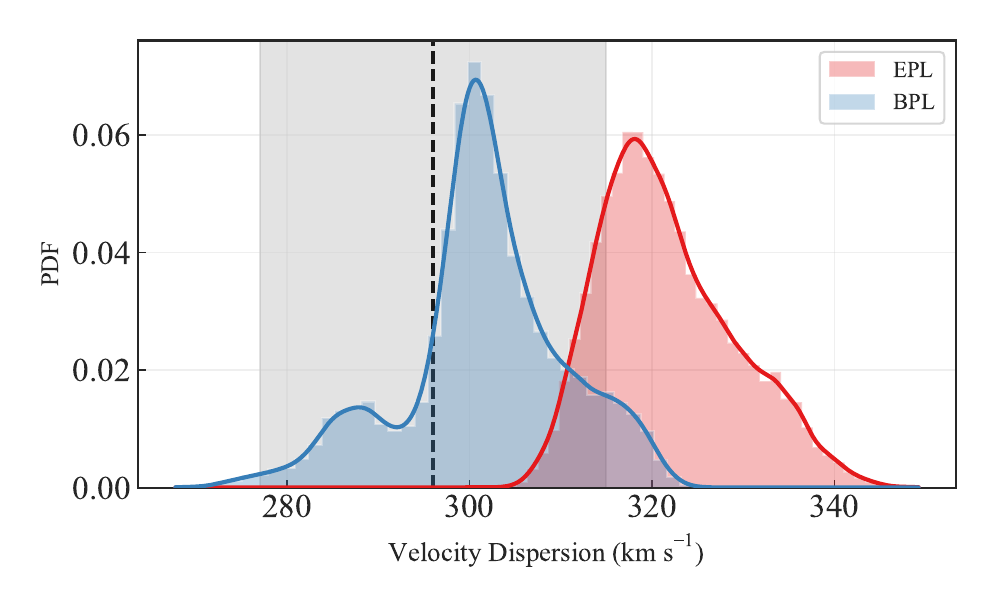}
\caption{Posterior distributions of the stellar velocity dispersion without applying the \((1-\kappa_{\rm ext})\lambda_{\rm int}\) correction. The grey shaded band and dashed line mark the observed aperture velocity dispersion, \(\sigma_{\rm ap}=296\pm19\,\rm km\,s^{-1}\).}
\label{fig:vel_dispersion_no_lambda}
\end{figure}

\begin{figure}
\centering
\includegraphics[width=\columnwidth]{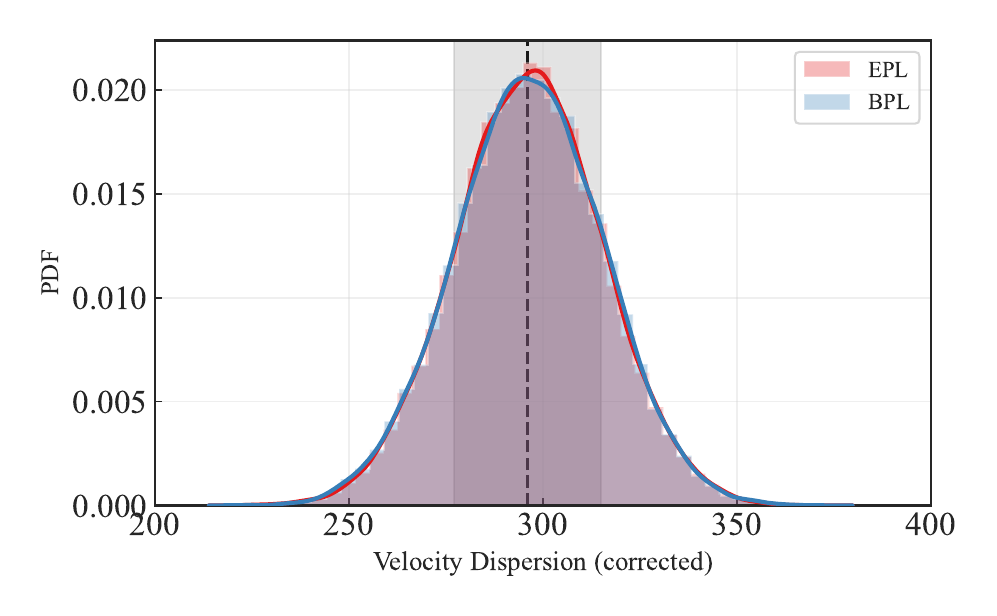}
\caption{Posterior distributions of the stellar velocity dispersion after applying the full \((1-\kappa_{\rm ext})\lambda_{\rm int}\) correction. The grey shaded band and dashed line mark the observed aperture velocity dispersion, \(\sigma_{\rm ap}=296\pm19\,\rm km\,s^{-1}\).}
\label{fig:vel_dispersion_with_lambda}
\end{figure}

We additionally compare the posterior distributions of the stellar velocity dispersion predicted by the two model families. The velocity-dispersion posteriors shown in Fig.~\ref{fig:vel_dispersion_no_lambda} are computed directly from the lens-model parameters using Eq.~(\ref{eq:sigmav_model}). To account for the full mass-sheet transformation, we then apply the $(1-\kappa_{\rm ext})\lambda_{\rm int}$ correction as prescribed by Eq.~(\ref{eq:sigmav_combine}), and the resulting corrected posteriors are shown in Fig.~\ref{fig:vel_dispersion_with_lambda}. These figures illustrate that, when the full mass-sheet correction is applied and $\lambda_{\rm int}$ is allowed to vary, both model families can formally reproduce the observed kinematic constraint. We stress, however, that this agreement should not be over-interpreted as an independent validation of the models, since $\lambda_{\rm int}$ can absorb residual mismatches associated with the adopted mass model and the inferred line-of-sight convergence. The differences in the posterior shapes therefore mainly reflect how the assumed mass profile couples to the kinematic information, $\lambda_{\rm int}$, and the line-of-sight convergence within the joint inference framework.

\begin{figure}
\centering
\includegraphics[width=\columnwidth]{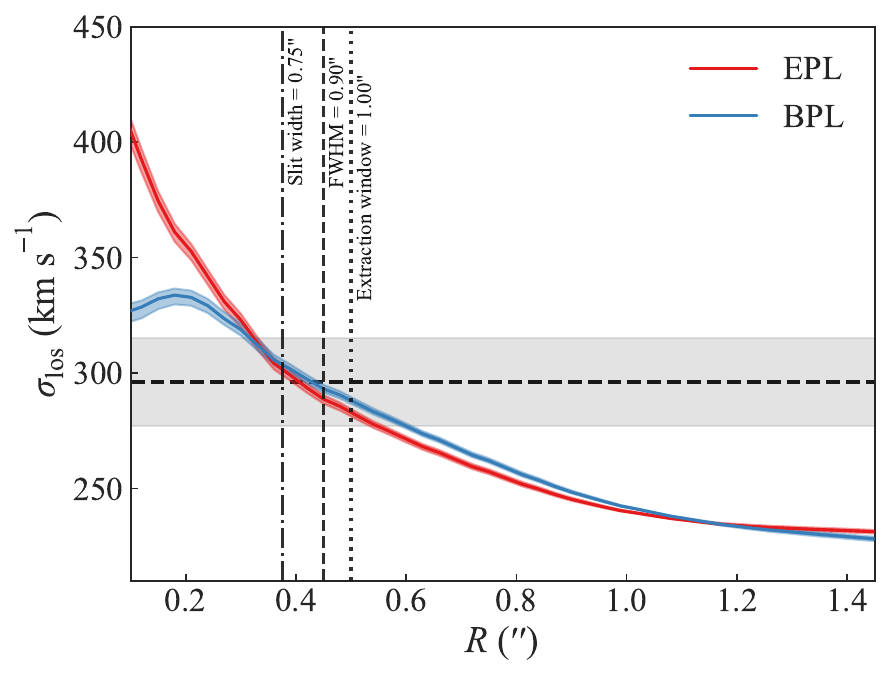}
\caption{The circularized line-of-sight velocity-dispersion profile, \(\sigma_{\rm los}(R)\), is shown as a function of projected radius \(R\), where the red and blue curves correspond to the EPL and BPL models, respectively. The solid lines and shaded bands denote the posterior median and the \(1\sigma\) credible range. We use the posterior samples from the BIC-selected best model set and predict the quantity measured in the IFU annular bins, assuming isotropic stellar orbits and a negligible PSF. The BPL model exhibits a noticeably flatter central velocity dispersion profile and shows a slight decrease near the innermost radius. The grey horizontal band and dashed line indicate the measured aperture velocity dispersion, \(\sigma_{\rm ap}=296\pm19\,\rm km\,s^{-1}\). The three vertical lines mark \(R=0\farcs375\), \(0\farcs45\), and \(0\farcs50\), corresponding to half the slit width (\(0\farcs75/2\)), half the seeing FWHM (\(0\farcs90/2\)), and half the extraction-window length (\(1\farcs00/2\)), respectively, for the Gemini/GMOS-S velocity-dispersion measurement. The BPL model exhibits a noticeably flatter central velocity dispersion profile and shows a slight decrease near the innermost radius.}
\label{fig:vdp}
\end{figure}

For further illustration, we also plot the circularized LOS velocity dispersion profiles, \(\sigma_{\rm los}(R)\), implied by the best model set (with minimal BIC value) for each family, shown in Fig.~\ref{fig:vdp}. Compared to the more sharply peaked central prediction of the EPL model, the BPL model yields a noticeably flatter inner profile within \(R \lesssim 0\farcs25\) and even shows a mild downturn at the smallest radii. Interestingly, spatially resolved observations of nearby massive early-type galaxies do include cases with an approximately flat central velocity dispersion followed by a decline at larger radii \citep{Veale2018}. At \(R \gtrsim 0\farcs35\), however, the two profiles nearly overlap and are in excellent agreement. Nonetheless, the innermost \(\lesssim 0\farcs2\) region is particularly susceptible to PSF and pixel size, and these profiles primarily to highlight potential differences in the predicted kinematic behaviour under the model degeneracies.

The comparisons presented in this section are intended to illustrate the internal consistency of the modelling and to provide physical intuition for how mass-profile flexibility influences lensing and kinematic observables. They support the interpretation adopted in the main text that the differences between the EPL and BPL results arise from model-dependent assumptions about the mass distribution, rather than from artefacts of light-profile modelling, PSF treatment, or numerical implementation.

\section{Conclusions}
\label{sec:conclusions}

In this work, we have developed and implemented a Broken Power Law (BPL) mass profile within the \textsc{Lenstronomy} framework, together with a numerical solver for the lensing potential and time delays. We validated the numerical implementation in the limit of EPL where we have the analytical expression. The results demonstrate excellent agreement, confirming the high accuracy of the numerical method.
We then applied the BPL model to a realistic lensed quasar system, namely WGD~2038--4008, incorporating multi-band imaging data, stellar kinematics, and LOS environmental information. For comparison, we also apply the EPL mass model to the same system, reproducing the result of \citet{Shajib2022}.

Both the EPL and BPL models are capable of reproducing the observed lensed morphology and image configurations with comparable quality, indicating that the current imaging data alone are not able to uniquely determine the inner mass profile of the lens galaxy. In the absence of a clearly detected central image, lensing image data could merely constrain the projected mass distribution near the Einstein radius, while providing weak sensitivity to the mass profile at radii well inside this scale. When stellar kinematics and LOS external convergence are incorporated to constrain the mass-sheet degeneracy, the two model families yield time-delay distances that differ at the \(\sim 20\)–25\% level, with \(D_{\Delta t}^{\rm EPL} = 1.61^{+0.44}_{-0.39}\,\mathrm{Gpc}\) and \(D_{\Delta t}^{\rm BPL} = 1.30^{+0.36}_{-0.32}\,\mathrm{Gpc}\). These differences propagate into the inferred Hubble constant, leading to \(H_0^{\rm EPL} = 61.05^{+19.14}_{-13.14}\,\mathrm{km\,s^{-1}\,Mpc^{-1}}\) and \(H_0^{\rm BPL} = 75.17^{+23.02}_{-16.28}\,\mathrm{km\,s^{-1}\,Mpc^{-1}}\) under a flat \(\Lambda\)CDM cosmology with fixed \(\Omega_{\rm m}=0.3\). For comparison, in the diagnostic case with \(\lambda_{\mathrm{int}}=1\) (Appendix~\ref{app:lambda1}), the Hubble constants are \(H_0^{\rm EPL} = 66.09^{+18.83}_{-12.81}\,\mathrm{km\,s^{-1}\,Mpc^{-1}}\) and \(H_0^{\rm BPL} = 74.24^{+20.29}_{-13.81}\,\mathrm{km\,s^{-1}\,Mpc^{-1}}\).

Importantly, the dominant contribution to this shift in \(H_0\) does not arise from a direct measurement of the innermost mass distribution, but rather from differences in the effective mass profile in the vicinity of the Einstein radius once the internal mass-sheet factor is constrained by stellar kinematics. The additional radial flexibility of the BPL parameterization allows for a redistribution of mass at small radii that remains consistent with the imaging constraints near the Einstein scale, and this redistribution propagates through the mass-sheet degeneracy into the inferred time-delay distance. For a single lens system, the resulting uncertainties remain substantial; nevertheless, this comparison illustrates that assumptions about the lens mass profile alone can induce systematic variations in inferred cosmological parameters at a level that is non-negligible in the context of current precision goals in time-delay cosmography.

The combined impact of internal mass-profile flexibility and LOS convergence can be summarized through the effective mass-sheet factor \(\Lambda_{\rm tot} = (1-\kappa_{\rm ext})\lambda_{\rm int}\). The inferred \(\Lambda_{\rm tot}\) distributions differ between the EPL and BPL families, reflecting how internal profile assumptions and environmental convergence trade off along the mass-sheet degeneracy. As emphasized throughout this work, both \(\lambda_{\rm int}\) and \(\kappa_{\rm ext}\) are inferred quantities that remain conditional on the adopted mass model. The resulting \(\Lambda_{\rm tot}\) should therefore be interpreted as the outcome of a joint, model-dependent inference rather than as an independent criterion for selecting a preferred physical description of the lens galaxy.

Comparisons presented in the Sect.~\ref{app:psidv}, including diagnostics of the Fermat potential, deflection angles, stellar velocity dispersion, and PSF residuals (Appendix~\ref{app:psf}), further support the internal consistency of the modeling and provide physical intuition for the origin of these model-dependent trends, without independently discriminating between the two mass-profile families.

Our results are broadly consistent with previous strong-lensing analyses of WGD~2038--4008 that employed different modeling frameworks and assumptions \citep[e.g.,][]{Shajib2019,Shajib2022}, and they echo earlier findings that time-delay measurements of individual lens systems can exhibit non-negligible sensitivity to mass-profile systematics \citep[e.g.,][]{Schneider2013,2021A&A...652A...7C,Blum2020}. While such effects do not allow one to draw conclusions regarding the resolution of the Hubble tension, they underscore the importance of flexible mass modeling and explicit marginalization over profile uncertainties in precision time-delay cosmography.

Overall, this study highlights that internal mass-profile systematics constitute a significant source of uncertainty in time-delay cosmography for individual lenses. Flexible mass models, combined with a consistent treatment of stellar kinematics and LOS effects, are therefore essential for robust cosmological inference, particularly in future hierarchical analyses that combine multiple lens systems within a unified modeling framework.

\section*{Acknowledgements}
We thank the anonymous referee for their insightful comments, constructive suggestions, and for pointing out issues that improved this manuscript. We thank Tian Li for valuable discussions and for sharing insights on numerical strategies for computing the Fermat potential integral. We are grateful to Simon Birrer and Anowar J. Shajib for making the \textsc{lenstronomy}-based modelling tools publicly available. In particular, we thank Anowar J. Shajib for helpful correspondence on the modelling of WGD~2038--4008. We also thank Alessandro Sonnenfeld for insightful comments. This work is supported in part by the National Natural Science Foundation of China Grants No. 12333001, No. 12541301 and No. 12541302 and by the China Manned Spaced (CMS) program with grant Nos. CMS-CSST-2025-A02, CMS-CSST-2025-A03, CMS-CSST-2025-A04 and CMS-CSST-2025-A20.

\section*{Data Availability}

The lens-modelling pipeline used in this work is built upon the publicly available software \textsc{Lenstronomy} \citep{Birrer2018}. 
Readers can access the code at \url{https://github.com/lenstronomy/lenstronomy}.
The Broken Power-Law (BPL) mass profile used in this work has been implemented by us and has now been incorporated into the main \textsc{Lenstronomy} code base, making it publicly available through the official repository.

The HST imaging data for the quadruply imaged quasar WGD~2038--4008 analysed in this paper are available from the Mikulski Archive for Space Telescopes (MAST) under programme HST-GO-15320. We follow the HST lens-processing workflow described by \citet{Tan2024} for data retrieval and preprocessing. 
In addition, we used the public DES Y3 redMaPPer cluster catalogue release \citep{Abbott2025} to check whether a massive nearby cluster is present along the line of sight to WGD~2038--4008.
The derived data products from our analysis (e.g.\ model posteriors, reconstructed source images, and related summary products) will be made available upon reasonable request to the corresponding author.

Our overall analysis closely follows the publicly released TDCOSMO modelling framework for WGD~2038--4008 \citep{Shajib2022} and is executed primarily in \textsc{Jupyter} notebooks \citep{Kluyver16}. For image download and preprocessing, we used \textsc{astroquery} \citep{Ginsburg2019} and \textsc{DrizzlePac} \citep{Fruchter2010}, together with \textsc{AstroObjectAnalyser}\footnote{\url{https://github.com/sibirrer/AstroObjectAnalyser}}. The lens modelling, posterior analysis, and figure production rely on \textsc{lenstronomy} \citep{Birrer2015,Birrer2018,Birrer21b} and the fast elliptical deflection-angle implementation \textsc{fastell4py}\footnote{\url{https://github.com/sibirrer/fastell4py}}, which wraps the original Fortran \textsc{fastell} code \citep{Barkana99}. Parameter inference and posterior visualisation make use of \textsc{emcee} \citep{Foreman-Mackey13}, \textsc{dynesty} \citep{Skilling2004,Speagle19}, and \textsc{getdist}\footnote{\url{https://github.com/cmbant/getdist}}. Additional analysis utilities include \textsc{Photutils} \citep{Bradley20}, \textsc{seaborn} \citep{Waskom2021}, and \textsc{colossus} \citep{Diemer18}.

Throughout the workflow we used standard scientific Python packages, including \textsc{numpy} \citep{Oliphant15}, \textsc{scipy} \citep{scipy2020}, \textsc{Astropy} \citep{AstropyCollaboration13,AstropyCollaboration18}, \textsc{SExtractor} \citep{Bertin96}, \textsc{pandas} \citep{mckinney2010}, and \textsc{matplotlib} \citep{Hunter07}. We further used \textsc{coloripy}\footnote{\url{https://github.com/ajshajib/coloripy/tree/master}} and \textsc{paperfig}\footnote{\url{https://github.com/ajshajib/paperfig}} for colour maps and publication-style figure utilities.


\bibliographystyle{mnras}
\bibliography{Time_Delay_BPL} 

\appendix

\section{Lens and Source Light Profiles}
\label{app:light}
In this study, we use a combination of parametric and non-parametric light profiles to model the lens galaxy and the source galaxy. The lens light is modeled using a combination of multiple elliptical \text{S\'{e}rsic} profiles, while the source light is modeled using a \text{S\'{e}rsic} profile for the bulge component and a shapelet expansion for the extended light distribution. The details of these profiles are as follows:

\begin{itemize}
  \item \textbf{\text{S\'{e}rsic} profile:} For the lens light in each band, we model the light distribution using a series of elliptical \text{S\'{e}rsic} profiles. The \text{S\'{e}rsic} profile is given by \citep{Sersic1963,Sersic1968}:
  \begin{equation}
    I_{\text{S\'{e}rsic}}(r) = I_0 \exp \left[ -b_n \left( \frac{r}{r_{\rm eff}} \right)^{1/n} \right]\;,
  \end{equation}
  where \( I_0 \) is the central intensity, \( r_{\rm eff} \) is the effective radius, \( n \) is the \text{S\'{e}rsic} index, and \( b_n \) is a constant that ensures that the effective radius encloses half the total flux. For the lens light, we use a combination of three elliptical \text{S\'{e}rsic} profiles for different bands, which adequately capture the lens galaxy's light distribution.
  
\item \textbf{Shapelets profile:} Shapelets provide a convenient basis to represent the surface brightness distribution of an extended source as a truncated linear expansion in localized, orthonormal functions. The Shapelets profile is given by \citep{Refregier2003,Refregier2003b}:
\begin{equation}
I(x,y) \;=\; \sum_{n_1+n_2 \le n_{\rm max}} f_{n_1,n_2}\,\psi_{n_1,n_2}(x,y;\beta),
\end{equation}
where \((x,y)\) are coordinates, \(\beta\) sets the characteristic spatial scale of the basis functions, \(f_{n_1,n_2}\) are the shapelet coefficients, and the expansion is truncated at total order \(n_{\rm max}\). The number of coefficients in this truncation is
\begin{equation}
N_{\rm shapelets} \;=\; \frac{(n_{\rm max}+1)(n_{\rm max}+2)}{2},
\end{equation}
corresponding to all integer pairs \((n_1,n_2)\) with \(n_1\ge 0\), \(n_2\ge 0\), and \(n_1+n_2 \le n_{\rm max}\). In the Cartesian formulation, the 1D normalized Hermite functions are defined as
\begin{equation}
\phi_n(x) \;\equiv\; \left(2^n \sqrt{\pi}\,n!\right)^{-1/2}\,H_n(x)\,\exp\!\left(-\frac{x^2}{2}\right),
\end{equation}
where \(H_n(x)\) is the Hermite polynomial of order \(n\) and \(x\) is a dimensionless coordinate. The 2D shapelet basis function is then given by the separable product
\begin{equation}
\psi_{n_1,n_2}(x,y;\beta) \;\equiv\; \phi_{n_1}\!\left(\frac{x}{\beta}\right)\,\phi_{n_2}\!\left(\frac{y}{\beta}\right).
\end{equation}

  \item \textbf{PL-\text{S\'{e}rsic} profile:} In addition to the \text{S\'{e}rsic} and shapelets profiles, we also experimented with a hybrid “PL–\text{S\'{e}rsic}” profile for the lens light, which combines a power-law mass profile with a \text{S\'{e}rsic} light profile. The formula for the 3D PL-\text{S\'{e}rsic} profile is \citep{TerzicGraham2005}:
\begin{equation}
I(r) = 
\begin{cases}
    j_c \left( \frac{r}{r_c} \right)^{-\alpha_c}  & \text{if } r \le r_c, \\
    j_0 \left( \frac{r}{s} \right)^{-u} \exp\left[ -\left( \frac{r}{s} \right)^\nu \right] & \text{if } r \ge r_c,
\end{cases}
\end{equation}
where
\begin{equation}
j_0 = j_c \left( \frac{r_c}{s} \right)^u \exp\left[ \left( \frac{r_c}{s} \right)^\nu \right],
\end{equation}
where \( r_c \) is the break radius, \(j_c\) is the luminosity density at \( r_c \), and \(s = \frac{R_{\text{eff}}}{k^n}\), where \(k\) is a function of the S\'{e}rsic index \(n\) \citep{Ciotti1999,MacArthur2003}. \(\nu = 1/n\) and  \(u \;=\; 1 - 0.6097\,\nu + 0.054635\,\nu^2\), following \citet{LimaNeto1999,Marquez2001}. This profile is intended to combine the effects of the mass distribution's power-law (in the inner regions) and the \text{S\'{e}rsic} profile (in the outer regions). However, this profile did not perform as well in initial tests, likely due to insufficient code maturity or prior settings. We plan to revisit and improve the PL-\text{S\'{e}rsic} profile in future work, focusing on enhancing the fitting process and exploring its potential advantages for complex galaxy structures.
\end{itemize}

\section{PSF Comparison Across EPL and BPL Models}
\label{app:psf}
\begin{figure}
\centering
\includegraphics[width=\columnwidth]{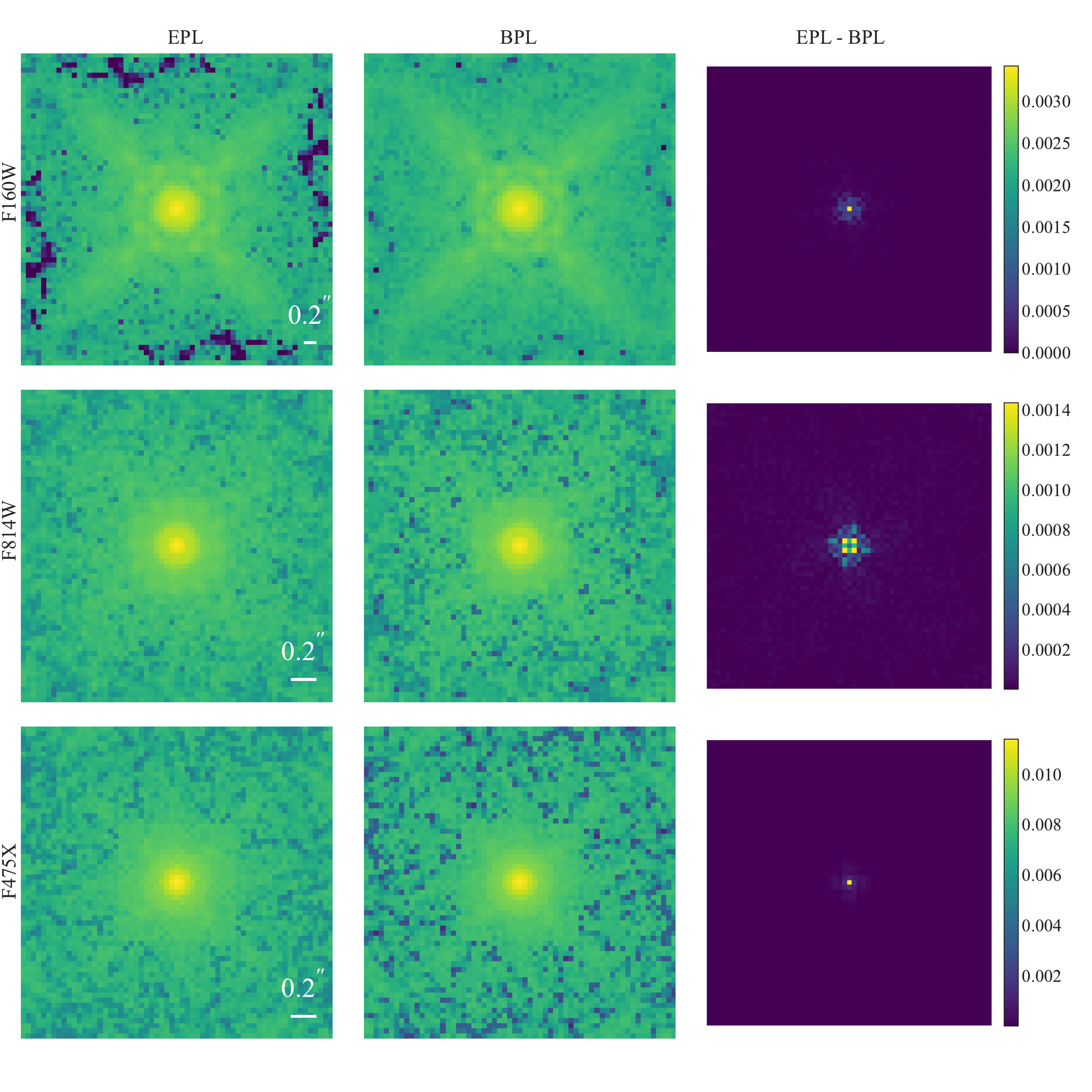}
\caption{PSF comparison between the EPL (left column) and BPL (middle column) families across three bands. The third column shows the residuals obtained by subtracting the BPL PSF from the EPL PSF. The comparison highlights minor differences in the PSF, with the BPL model providing a slightly better fit to the observed data, particularly in the central regions.}
\label{fig:psf_comparison}
\end{figure}

To assess the effect of different lens models on the PSF and residuals, we compare the PSF-convolved lens models (EPL vs. BPL) across the three imaging bands. The first column of Fig.~\ref{fig:psf_comparison} shows the PSF for the EPL family, the second column shows the corresponding PSF for the BPL family, and the third column shows the residuals obtained by subtracting the BPL PSF from the EPL PSF. These residuals provide a direct comparison between the two models.

The differences between the two models are minimal, indicating that both reproduce the PSF structure with comparable accuracy across all bands. As seen in the residual maps, the BPL model provides only a marginally better match in the central regions, but the improvement is negligible within the noise level. Overall, the PSF reconstructions are virtually indistinguishable, and the FWHM of the PSF is identical between the two models at the corresponding pixel scale. 
We also note that the reconstructed PSF still shows a certain level of noise. In future work, we will explore PSF denoising strategies to further reduce this source of uncertainty in the lens modelling.

\label{lastpage}
\section{Case with fixed internal mass-sheet factor}
\label{app:lambda1}

\begin{figure}
  \centering
  \includegraphics[width=\columnwidth]{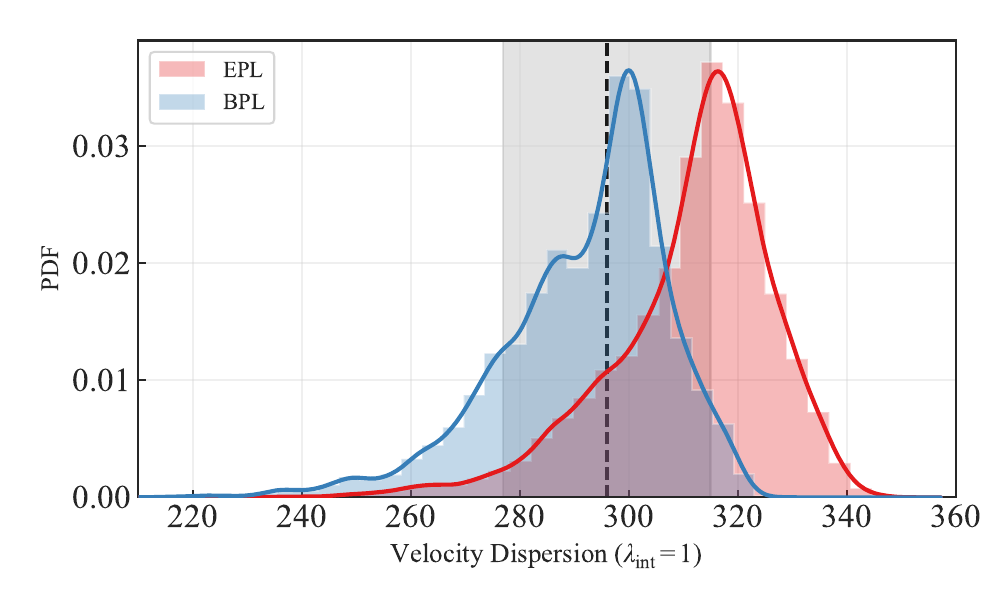}
  \caption{Posterior distributions of the stellar velocity dispersion for the diagnostic case \(\lambda_{\rm int}=1\), i.e. after applying only the LOS correction \((1-\kappa_{\rm ext})\). The grey shaded band and dashed line indicate the observed aperture velocity dispersion, \(\sigma_{\rm ap}=296\pm19\,\rm km\,s^{-1}\). In this case, the posteriors of both the BPL and EPL models are shifted toward the observed velocity-dispersion constraint.}
  \label{fig:vel_dispersion_lambda1}
\end{figure}

For the case with \(\lambda_{\rm int}=1\), we fix the internal mass-sheet factor and therefore apply only the LOS correction \((1-\kappa_{\rm ext})\) to the lensing and kinematic quantities. This case provides a useful controlled comparison, because it removes the additional freedom associated with the internal MST and isolates the part of the model dependence that is already present in the underlying EPL and BPL mass-profile families \citep{Schneider2013,Birrer2020,Shajib2022}. The corresponding velocity-dispersion posteriors are shown in Fig.~\ref{fig:vel_dispersion_lambda1}. In this case, once \(\lambda_{\rm int}\) is fixed to unity, the stellar kinematics can no longer enforce an exact sample-by-sample re-scaling of the model prediction. Instead, the kinematic likelihood acts through importance weighting, preferentially selecting samples whose LOS-corrected velocity dispersions are closer to the measured aperture value. As a result, the posteriors of both the BPL and EPL models are shifted toward the observed velocity-dispersion constraint, while the remaining difference in their peak locations and overall shapes reflects how the two lens-model families differ in their radial mass structure after the LOS correction.

\begin{figure}
  \centering
  \includegraphics[width=\columnwidth]{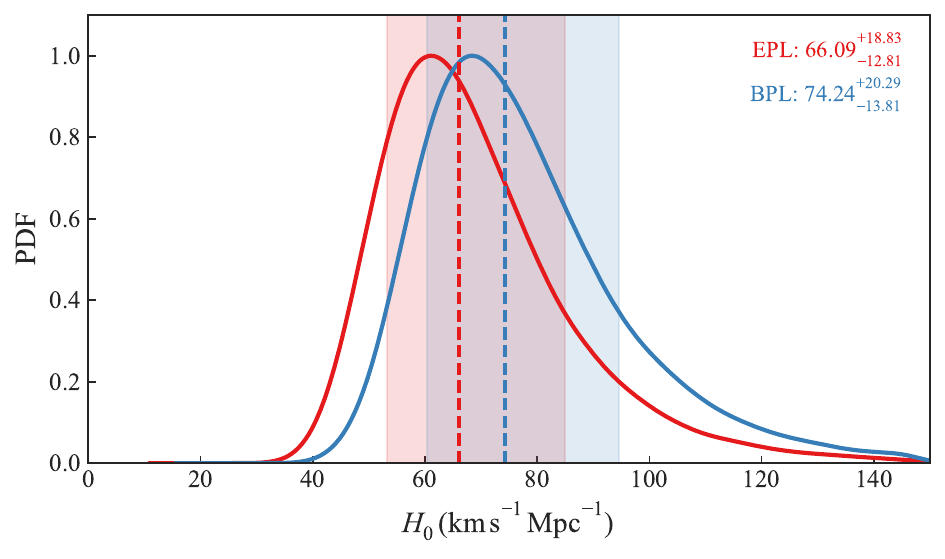}
  \caption{Posterior distributions of \(H_0\) obtained from the \(\lambda_{\rm int}=1\), assuming a flat \(\Lambda\)CDM cosmology with fixed \(\Omega_{\rm m}=0.3\).}
  \label{fig:h0_lambda1}
\end{figure}

The same trend propagates into the inferred Hubble constant. Using the corrected time-delay prediction together with the observed time delays of WGD~2038--4008 \citep{Wong2024}, we obtain \(H_0^{\rm EPL} = 66.09^{+18.83}_{-12.81}\,\mathrm{km\,s^{-1}\,Mpc^{-1}}\) and \(H_0^{\rm BPL} = 74.24^{+20.29}_{-13.81}\,\mathrm{km\,s^{-1}\,Mpc^{-1}}\) for the \(\lambda_{\rm int}=1\) case (Fig.~\ref{fig:h0_lambda1}). This shows that part of the larger model dependence in the full inference is driven by the extra internal MST freedom coupled to the stellar kinematics, while a non-negligible difference between the two model families is already present even when only the LOS correction is applied.

\end{document}